\pdfoutput=1

\documentclass{sig-alternate} 
\usepackage{mathptmx} 

\usepackage{graphicx}
\newcommand{\ignore}[1]{}
\usepackage[pass]{geometry}
\usepackage{fancyhdr}
\usepackage[normalem]{ulem}
\usepackage[hyphens]{url}

\usepackage[sort,nocompress]{cite}

\usepackage[final]{microtype}
\usepackage{flushend}
\usepackage[bookmarks=false]{hyperref}

\usepackage{alltt}
\usepackage{epsfig}
\usepackage{graphicx}
\usepackage{enumitem}
\usepackage{amsmath}

\usepackage{amssymb}
\usepackage[normalem]{ulem}
\usepackage{float}
\usepackage[]{algorithm2e}
\usepackage{lipsum}
\usepackage{balance}
\usepackage{tikz}
\usepackage{calc}
\usepackage{xcolor, colortbl}
\usepackage{multirow}
\usepackage{pifont}
\usepackage{adjustbox}
\usepackage{array}

\usepackage{tcolorbox}
\usepackage{amssymb,amsmath}

\usepackage{xfrac}
\usepackage{mdframed}

\usepackage{color,soul}
\definecolor{lightgreen}{RGB}{195, 233, 211}
\sethlcolor{lightgreen}

\soulregister\cite7
\soulregister\ref7
\soulregister\pageref7

\makeatletter
\def\SOUL@hlpreamble{%
	\setul{}{2.2ex}
	\let\SOUL@stcolor\SOUL@hlcolor
	\SOUL@stpreamble
}
\makeatother

\newcommand\thintilde{{\lower.92ex\hbox{\mathtt{\char`\~}}}}
\newcommand\thicktilde{{\lower.74ex\hbox{\texttt{\char`\~}}}}

\usepackage{amsfonts}
\usepackage{url}

\fancypagestyle{firstpage}{

	\fancyhf{}
	\setlength{\headheight}{50pt}
	
}

\usepackage{authblk}
\author[ ]{Vinson Young}
\author[ ]{Moinuddin K. Qureshi\vspace{-.1in}}
\affil[ ]{Georgia Institute of Technology}
\affil[ ]{\texttt {\{vyoung,moin\}@gatech.edu}\vspace{-.2in}}

\title{
\vspace{-0.2in}
\vspace{-0.48in}
To Update or Not To Update?:  Bandwidth-Efficient Intelligent Replacement Policies for DRAM Caches
\vspace{-0.48in}
\vspace{-.2in}
}

\begin{document}
\maketitle
\thispagestyle{firstpage}
\pagestyle{plain}

\ignore{
	Heterogeneous Memory Systems -> effective DRAM Cache designs help hide later level effects. Commodity DRAM Cache implement tags-in-ECC. Each way needs additional lookup, hence, minimally associative (one or two ways). We need good replacement policies tailored for low associativity.

	State-of-art Reuse-based Replacement (e.g. RRIP, Segmented-LRU, Multiperspective) policies protect lines that have had reuse, and installs other lines at low priority. However, such policies are ineffective for one-way (most degrade into Always-Install) or two-way (must still find victim even if both lines have predicted reuse) caches. Some policies provision for Probabilistic Bypass (BIP, CRC-1) to avoid evicting predicted-reuse lines, but these are coarse-grain bypass decisions that affect the whole cache instead of operating at a per-line basis. In this work, we advocate unifying the approaches to achieve per-line Reuse-based Bypassing. Improves hit-rate to improve 1-way cache perf by 10\%.

	Need bandwidth to promote on hit, demote/age on victim selection. Use spatial hints to reduce aging costs to trim cost to achieve 17\% perf.
}

\begin{abstract}

This paper investigates intelligent replacement policies for improving the hit-rate of gigascale DRAM caches. Cache replacement policies are commonly used to improve the hit-rate of on-chip caches. The most effective replacement policies often require the cache to track and update per-line reuse state to inform their decision. A fundamental challenge on DRAM caches, however, is that stateful policies would require significant bandwidth to maintain per-line DRAM cache state. As such, DRAM cache replacement policies have primarily been stateless policies, such as always-install or probabilistic bypass. Unfortunately, we find that stateless policies are often too coarse-grain and become ineffective at the size and associativity of DRAM caches. Ideally, we want a replacement policy that can obtain the hit-rate benefits of stateful replacement policies, but keep the bandwidth-efficiency of stateless policies.

We perform our study on a DRAM cache design similar to the one used in Knights Landing, and find that tracking per-line reuse state can enable an effective replacement policy that can mitigate the common thrashing patterns seen in gigascale caches. We propose a stateful replacement/bypass policy called \textit{RRIP Age-On-Bypass (RRIP-AOB)}, that tracks reuse state for high-reuse lines, protects such lines by bypassing other lines, and \textit{\underline{A}ges the state \underline{O}n cache \underline{B}ypass}. Unfortunately, such a stateful technique requires significant bandwidth to update state. To this end, we propose \textit{Efficient Tracking of Reuse (ETR)}. ETR makes state tracking efficient by accurately tracking the state of only one line from a region, and using the state of that line to guide the replacement decisions for other lines in that region. ETR reduces the bandwidth for tracking the replacement state by 70\%, and makes stateful policies practical for DRAM caches. Our evaluations with a 2GB DRAM cache, show that our RRIP-AOB and ETR techniques provide 18\% speedup while needing less than 1KB of SRAM. 


\end{abstract}






\ignore{

Effective DRAM cache designs enable a heterogeneous memory system that can simultaneously provide both high bandwidth (of 3D-stacked DRAM) and high capacity (of commodity DRAM or 3D XPoint).
However, effective DRAM cache designs are difficult to design as both the DRAM and interconnect technology are highly sensitive to bandwidth consumption -- traditionally effective cache-optimizations, such as associativity, often become ineffective in DRAM caches due to high bandwidth consumption. 
In this paper, we investigate cache replacement policies in the context of bandwidth-sensitive commercial DRAM caches. We find that an effective replacement policy must not only improve cache hit-rate, but also keep its bandwidth consumption (for updating replacement state in DRAM) in check.

We build upon the direct-mapped, tags-in-ECC DRAM cache design in Knights Landing product, as it is the only commercially available DRAM cache design for CPUs.
However, such caches have limited associativity (1 to 2 ways) and current replacement policies do not scale to such associativities.
To enable intelligent reuse-based replacement policies for such commercial DRAM caches with limited associativity, this paper proposes a bypass version of RRIP \textit{(RRIP-BYP)}, whereby if a good victim cannot be found, the incoming line is bypassed and the reuse counters are aged.  We also propose \textit{Efficient Tracking of Reuse (ETR)} to avoid the bandwidth overheads of updating reuse counters. ETR makes state tracking efficient by accurately tracking the state of only one line from a region, and using the state of that line to guide the replacement decisions for other lines in that region. ETR reduces the bandwidth for tracking the replacement state by 70\%. Our evaluations with a 2GB direct-mapped DRAM-cache, show that our proposal provides an average speedup of 18.0\% while needing an SRAM overhead of less than 1KB.
We also show that our solution extends to two-way designs, and can be used with other replacement policies (SHiP).







}

\ignore{
------------------------
HPCA version

Practical DRAM caches, such as the one in Intel Knights Landing (KNL), are designed to be direct-mapped. The hit rate of such cache designs can be improved using replacement policies that decide between either evicting the line resident in the cache or bypassing the incoming line. Unfortunately, prior works on DRAM cache bypassing has been limited to probabilistic bypass policies that do not exploit the variability in reuse behavior of individual lines, and hence have limited speedup (3\%). In this paper, we investigate how state-of-the-art policies (such as RRIP or SHiP) that exploit per-line reuse information, can be applied to practical DRAM caches. 

There are two obstacles in using reuse-based replacement policies in practical DRAM caches: (1) These existing reuse-based policies were designed to operate by comparing multiple reuse counters within the set -- they  become ill-defined when the set contains only one counter, therefore these policies cannot be directly applied to direct-mapped designs (2) Even if the policies could be extended to DRAM caches, storing the per-line counter incurs either impractical storage (8MB SRAM) or consumes significant bandwidth for performing counter updates, if counters are stored in DRAM.

To enable reuse-based replacement policies for practical DRAM caches, this paper proposes {\em Direct-Mapped Edition (DME)} of RRIP (RRIP-DME) whereby the event of bypassing of the incoming line is used to update the reuse counter of the resident line.  We also propose \textit{Efficient Tracking of Reuse (ETR)} to avoid the bandwidth overheads of updating the reuse counters. ETR makes state tracking efficient by accurately tracking the state of only one line from a region, and using the state of that line to guide the replacement decisions for other lines in that region. ETR reduces the bandwidth for tracking the replacement state by 70\%. Our evaluations with a 2GB DRAM-cache, show that our proposal provides an average speedup of 18.0\% while incurring an SRAM overhead of less than 1KB. Our design performs within 2\% of an idealized design that has 8MB of SRAM for maintaining reuse counters. We also show that our solution can be used with other replacement policies (SHiP) and significantly outperforms the recently proposed ACCORD design (10\% versus 18\%).
}

\ignore{
------------------------
old version

This paper investigates intelligent replacement policies for commercially-available DRAM cache designs, such as those implemented in Intel Knights Landing. Today's DRAM caches use a direct-mapped design, co-locate the tag and data within the DRAM array, and stream out the tag and the data concurrently on an access.  Unfortunately, such a direct-mapped design can have low hit-rate.  
Intelligent replacement policies (e.g., RRIP) typically rely on tracking reuse, and the reuse state gets updated on both cache-hits and cache-bypasses. Updating the replacement state consumes significant bandwidth if the  state is kept in DRAM and keeping the replacement state in SRAM would incur  unacceptably large storage overheads. Ideally, we want to keep the replacement state-in-DRAM (SID) and yet obtain the performance of a state-in-SRAM (SIS) design. 

This paper proposes \textit{Efficient Tracking of Reuse (ETR)} to enable intelligent replacement policies for practical DRAM caches. ETR is based on the key observation that multiple lines from a page that are coresident in the DRAM cache at a given time tend to have similar replacement state. ETR makes state tracking efficient by accurately tracking the state of only one of the coresident lines from the page and using the state of that line to guide the replacement decisions for other coresident lines of that page. ETR reduces the bandwidth for tracking the replacement state by 70\%. We discuss implementation of ETR with signature-based policies. We also compare to associative designs. Our evaluations with a 2GB DRAM-cache, show that RRIP-DME + ETR provides a speedup of 18.0\% while incurring an SRAM overhead of less than 1KB.  ETR performs within 2\% of an idealized SIS design that would need 8MB of SRAM.
}

\ignore{

in the DRAM array results in su

line must be byassed or the restreams out the tag and the data  places tags with the line.

-locates the tags within the ECC bits of the line.  A direct-mapped cache can have low-hit rate 

Current DRAM-caches are \textit{direct-mapped}, with \textit{tags stored in ECC bits}, to handle the issue of tag-storage and tag-access. Such a direct-mapped design is effective for reducing access-latency and bandwidth consumption. However, this direct-mapped organization comes at a cost to hit-rate. 
In this work, we target improving DRAM-cache hit-rate.

A traditional method for improving cache hit-rate is using effective replacement policies. However, intelligent replacement policies typically require remembering reuse. We can store reuse and replacement state alongside tags in the DRAM-ECC space. But, maintaining this Replacement-State in DRAM (RSD) incurs significant bandwidth overhead. For example, Re-Reference Interval Predictor requires \textit{promote} bandwidth on hit, and \textit{demote} bandwidth when selecting victims. If we formulate RRIP as a bypassing policy, these bandwidth overheads limit the performance potential of the RSD-RRIP approach to 10\% speedup. We can store Replacement-State in SRAM (RSS) to avoid these bandwidth costs and achieve 21\% speedup. However, storing replacement state for gigascale caches on-chip requires large multi-MegaByte SRAM storage (8MB).

Current DRAM-caches are direct-mapped to handle the issue of tag-storage and tag-access. While a direct-mapped design is effective for reducing access-latency and bandwidth consumption, this direct-mapped organization comes at a cost to hit-rate. 
In this work, we target improving DRAM-cache hit-rate and bandwidth through intelligent replacement policies. However, storing replacement state for gigascale caches on-chip requires impractically large multi-MegaByte SRAM storage.

We can avoid the significant state-in-SRAM storage overhead by in-lining replacement bits alongside tags in DRAM. However, such a state-in-DRAM approach requires significant bandwidth overhead to update state. For instance, intelligent replacement policies such as RRIP often require state-update on hit, and state-update on victim-selection. This actually causes increased DRAM-cache bandwidth consumption on average. The state update bandwidth costs limit the speedup the state-in-DRAM approach can offer to 10\%, as opposed to the 21\% an impractical state-in-SRAM approach can offer. We need effective mechanisms to reduce the state-update cost to make state-in-DRAM perform well.

To tackle storage and bandwidth costs, we implement a Trimmed RRIP by storing state-in-DRAM and reducing state update bandwidth by coordinating replacement policy across sets. Coordinating replacement policy reduces 70\% of the update costs as we now need to maintain replacement information for just the first conflicting set of a page. Additional modifications for PC and write-awareness enable further state-update reduction to 80\%. Our bandwidth-efficient intelligent replacement policy enables 21\% speedup across a wide range of workloads, while requiring less than 2 Kilobytes of SRAM storage.
}

\ignore{
Practical DRAM-caches in the near future are likely to be direct-mapped or minimally-associative to handle the issue of tag-storage and tag-access. While a direct-mapped design is effective for reducing access-latency and bandwidth consumption, this direct-mapped organization comes at a cost to hit-rate. 
Throughout this work, we target improving DRAM-cache hit-rate and bandwidth through intelligent replacement or bypassing policies. 

A primary method for improving cache hit-rate is implementing effective replacement policies. 
Our insight is that we can utilize the concept of cache bypassing to implement replacement policies.
By viewing bypassing as a second way, we can formulate intelligent replacement polices as DRAM-cache bypassing policies and exploit recent insights to improve DRAM-cache hit-rate.

Extensive studies in last-level caches have shown that reuse-aware intelligent replacement policies (RRIP, SHiP) can provide substantial hit-rate improvement. 
However, such intelligent replacement policies require capturing re-use behavior, which needs to update state on hit. 
On DRAM-caches, remembering such state incurs either multi-mega-byte SRAM-storage to maintain state-in-SRAM, or significant DRAM-bandwidth to maintain state-in-DRAM. 
To tackle storage and bandwidth costs, we implement RRIP by storing state in DRAM, and reduce state-update bandwidth by coordinating replacement policy across sets. This spatially-coordinated RRIP reduces 70\% of the state-update cost and achieves 18\% out of the 22\% speedup achievable by an impractical state-in-SRAM approach.
	
In addition, we further reduce state-update bandwidth by making the cache PC-aware with SHiP and write-aware with dynamic write-back / write-around policy. Adding SHiP enables further reducing state-update costs. And dynamic write-back / write-around is able to intelligently decide when to write-back (to reduce write costs) or write-around (to save cache capacity). In total, our bandwidth-efficient intelligent replacement policy reduces install-bandwidth by 85\% and reduces state-update bandwidth by 85\% (while costing less than 2KB SRAM storage) to achieve 21\% speedup, obtaining most of the 22\% speedup an impractical state-in-SRAM approach offers.
}

\ignore {

	Practical DRAM-caches in the near future are likely to be direct-mapped to handle the issue of tag-storage and tag-access. While a direct-mapped design is beneficial for access-latency and bandwidth, this direct-mapped organization can come at a cost to hit-rate. Tackling the problem of lowered hit-rate is paramount for scaling future applications where the working sets are growing to sizes similar or greater than DRAM-cache capacity. 
	
	A primary method for improving cache hit-rate is designing effective replacement policies that keep only the most useful lines in the cache. However, these designs have traditionally been tailored for caches that have many ways and many potential victims, and cannot be directly applied to our direct-mapped DRAM-caches. In this work, we ask if it is possible to distill the insights gained from prior work on cache replacement policies and similarly utilize them for improving hit-rate of direct-mapped DRAM caches.
	
	Through this work, we find that the concept of protecting working-set and bypassing less-useful lines can still be applied to a direct-mapped cache through the concept of DRAM-cache-bypassing. Instead of installing the line at a low priority, we can instead simply bypass (i.e. not-install) the line most of the time. This allows us to map many traditional replacement policies to DRAM-cache-bypassing policies and allows us to utilize the many insights gained from prior work. Throughout this work, we design and evaluate bypassing versions of BIP, DIP, TADIP, and current state-of-art SHiP++ on DRAM-caches, and find that they can improve system performance by -14.5\%, +4.1\%, +5.5\%, and +9.8\%, respectively, for PCM-based main memory.

	We also investigate a separate technique to \textit{reduce cache thrashing} which involves tailoring a thrash-resistant local replacement policy (RRIP) for DRAM-caches. RRIP is effective at improving hit-rate on a per-line basis by protecting re-used lines and bypassing incoming lines. However, it can come at a high cost for DRAM caches due to increased bandwidth consumption for state updates (i.e., promote on hit, and demote on bypass). We first design a bypassing version of RRIP for a direct-mapped cache. We find that an idealized RRIP with no state-update cost can achieve 22\% speedup. However, if we account for state-update bandwidth, the base RRIP only sees 8\% speedup. Through the combination of two state-update reducing techniques, a trimmed RRIP achieves 17\%, with no workloads experiencing degradation.
	
	We also give QoS robustness.



	

	


	
	
	

}

\ignore{
-- DRAM Cache DirectMapped KNL
-- Need Repl, Repl is bypassing
-- Random, Recency no good -- need reuse based repl (RRPV)
-- We extend RRPV and found that it improves hitrate and performance. But performance depends on where you store the state (SRAM costly, DRAM bandwidth)
-- The bandwidth required to do state update results in suboptimal performance, what if we coudl reduce this

-- Coresidency and Eviction Locality
-- ETR and how it works
-- DynWritePolicy (DWP)
-- We also show to implement more complex schemes such as SHiP with ETR 
-- Contribs

---------------------

ETR
1. We propose direct-mapped editions of RRIP suitable for DRAM caches.
2. We reduce the bandwidth needed for doing replacement state updates (needed when state is stored in DRAM).
3. We show these ideas are useful for SHiP as well as for the recent ACCORD proposal.

}

\vspace{.08in}
\section{Introduction}



DRAM caches are important for enabling effective heterogeneous memory systems that can transparently provide the bandwidth of high bandwidth memories~\cite{HBM}, and the capacity of high capacity memories~\cite{ddr4,intel:3dxpoint}.
Designs for DRAM cache organize the tag-store such that the tags can be kept in DRAM (to reduce storage overheads) and yet the tags can also be obtained with low latency and low bandwidth overheads~\cite{KNL,Alloy}. 
For example, Intel's Knights Landing product organizes its DRAM cache 
as a direct-mapped cache with tags stored alongside each data-line, so that one access can retrieve both tag and data. 
This direct-mapped design has been shown to be effective for enabling low latency and bandwidth-efficient tag access~\cite{Alloy}; however, such a direct-mapped design can have significant conflict misses. 
One could consider increasing associativity to improve hit-rate, but, increasing associativity also substantially increases bandwidth consumption and degrades performance for many workloads. Fortunately, cache bypassing~\cite{BEAR,ctrbypass,crc1} offers a way to both improve hit-rate and decrease bandwidth consumption, while still maintaining a direct-mapped organization.
We investigate the extent to which an intelligent bypass policy can reduce conflict misses for DRAM caches. We perform our evaluations on a direct-mapped DRAM cache similar to the one used in KNL~\cite{KNL,Alloy}.

We would like to use the most effective replacement policies to improve DRAM cache hit-rate.
However, intelligent replacement policies~\cite{RRIP,crc1,SHiP,SHiP++} often require the cache to track per-line state that needs to be updated on cache events. 
On a DRAM cache, managing this per-line state is difficult as tracking even 2 bits of state per line would require multi-megabyte storage. As such, DRAM cache designs would need to keep this state in the DRAM array, and spend offchip bandwidth to update state.
Prior replacement policies proposed for DRAM caches have avoided this per-line state with stateless policies~\cite{Alloy,KNL,BEAR}. The DRAM cache in KNL~\cite{Alloy,KNL}, for example, employs an \textit{Always-Install} policy.
Along the same lines, Chou et. al~\cite{BEAR} propose a policy that bypasses the cache with 90\% probability (we call this policy \textit{90\%-Bypass}).
However, such stateless policies often fail to capture the reuse patterns commonly seen in large caches. We show how such policies are often inadequate with an example.

\begin{figure*}[htb] 
	\centering
    \vspace{-0.25 in}
    \includegraphics[width=7.0in]{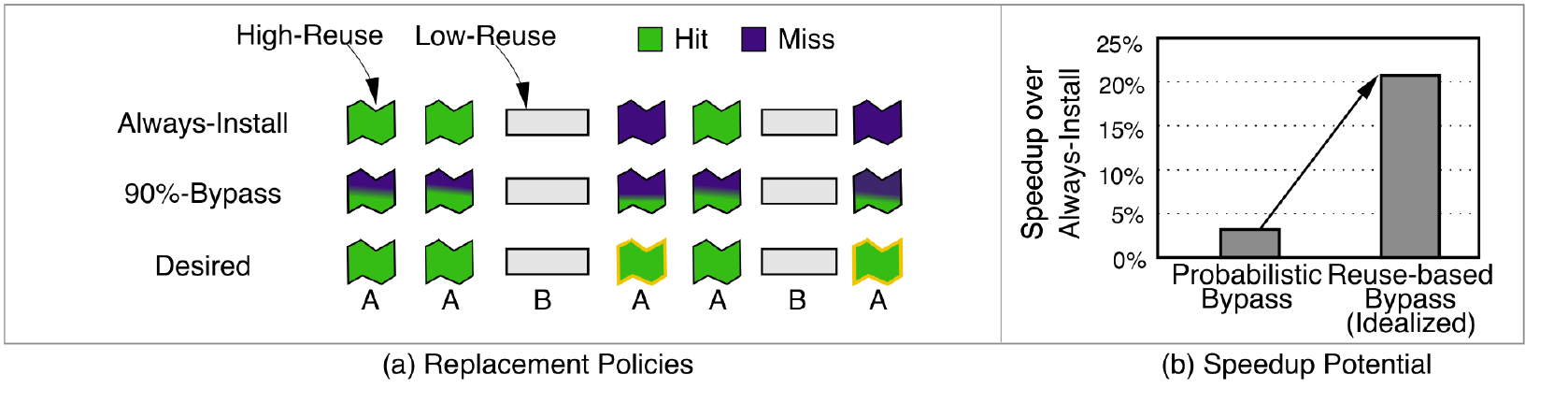}
    \vspace{-0.35 in}
    \caption{(a) Always-Install, 90\%-Bypass, and Desired replacement policies under mixed high-reuse low-reuse access pattern. (b) Potential for speedup: Probabilistic Bypass~\cite{BEAR}, and Ideal Reuse-based Bypass with no state update cost.}
    \vspace{-0.15 in}
	\label{fig:intro} 
\end{figure*}

Let us consider replacement policies for a common access pattern where the workload has repeated accesses to high-reuse data (labeled \textit{A}) interspersed with accesses to low-reuse data that is not re-referenced while it is in the cache (labeled \textit{B}), as shown in Figure~\ref{fig:intro}(a). For the baseline \textit{Always-Install} policy, accesses to A will install A and enable subsequent accesses to A to hit; however, accesses to low-reuse B will evict A and cause the subsequent access to A to miss. In this case, always-installing lines allows low-reuse B to evict high-reuse A, and this results in degraded hit-rate and wasted install bandwidth.
For a \textit{90\%-Bypass} policy, references to A will install some A lines and marginally improve hit-rate, and references to B will install only a few lines and marginally degrade hit-rate. In this case, 90\%-Bypass offers some working-set protection; however, it is indiscriminate in deciding which lines to protect and may not achieve high hit-rate. Figure~\ref{fig:intro}(b) shows that such a probabilistic bypass policy has poor performance potential of 3\%. Ideally, we desire a bypass policy that can remember and protect individual lines that have high reuse (i.e., A), and bypass other lines (i.e., B). 
Figure~\ref{fig:intro}(b) shows that if we are able to formulate such a \textit{reuse-based bypass} policy while avoiding the bandwidth cost for state update, we could achieve up to 20\% speedup.

Our approach to improving DRAM cache performance is to (1) design a \textit{reuse-based bypass} policy to improve DRAM cache hit-rate, and to (2) reduce the bandwidth cost of state update to further improve performance.

\newpage

In this paper, we use {\em Re-Reference Interval Prediction (RRIP)}~\cite{RRIP} as a representative example of a replacement policy that is designed to exploit reuse~\cite{RRIP,SHiP,crc1,Hawk-eye,multiperspective}.  RRIP requires that each line is equipped with metadata bits (two-bit counter called RRPV) to track reuse. RRPV is set to 0 on a hit, and the victim line is identified as a line that has an RRPV of 3. And, if no lines have an RRPV of 3, all counters in the set are incremented and victim-selection is repeated. 
While RRIP is effective for set-associative caches, it becomes ill-defined for direct-mapped caches, as such a cache would have only one counter in the set. 
Following the algorithm for selecting a victim will always cause the resident line to get evicted, even if the line had an RRPV of 0. Similarly, bypassing the incoming line if the resident line has an RRPV=0 will mean that such lines will never get evicted from the cache.


To enable reuse-based replacement policies for direct-mapped DRAM caches, we propose a bypass version of RRIP, which we call \textit{RRIP-AOB}. The key mechanism in RRIP-AOB, is to \textit{\underline{A}ge the counters \underline{O}n cache \underline{B}ypass}. For example, if a good victim cannot be found (no RRPV are 3), we bypass the incoming line and age the reuse counter (increment RRPV). After several bypass+age events, a resident line that is no longer useful will have its RRPV reach 3 and become a candidate for eviction. This enables the cache to protect lines that have had reuse via bypassing, but also provides a path to eventually victimize cold lines. Our insight makes RRIP (and other reuse-based policies) applicable to DRAM caches.

Another practical obstacle in implementing reuse-based policies for DRAM caches is the high state update cost of maintaining replacement state in DRAM.
A straight-forward way of implementing RRIP-AOB in DRAM cache is to extend the tag-entry of the line to incorporate the bits for tracking the replacement state of the line. However, it incurs bandwidth overhead for performing update of the replacement state: resetting the RRPV counter on a hit ({\em promotion}), and incrementing the RRPV counter on a bypassing miss ({\em demotion}). Note that these accesses for updating the replacement state are not present in the baseline and for designs that do bypassing without tracking per-line state.  
If we can completely remove with state update cost with Ideal RRIP-AOB, we can achieve up to 20\% speedup. To reduce the state update cost of maintaining per-line counters in DRAM, we propose \textit{Efficient Tracking of Reuse (ETR)}. 

ETR reduces the bandwidth consumed in performing updates of the replacement state by doing the updates for only a subset of the lines and using their replacement state to infer the replacement state of the other lines.  ETR is based on two key properties that we observe in DRAM caches:   {\em Coresidency} and {\em Eviction-Locality}.  Coresidency indicates that at any given time if a line is present, then several other line belonging to that 4KB region are also present in the cache. Eviction-Locality indicates that when a line gets evicted from the cache, the replacement-state of the other coresident lines belonging to that region tend to have similar replacement-state as the line being evicted. We show strong levels of coresidency and eviction-locality with RRIP-AOB. ETR exploits the properties of coresidency and eviction-locality to reduce the updates for tracking the replacement state. Rather than updating the replacement-state for all lines in the cache, ETR simply updates the replacement-state for one of the coresident lines of the region, and uses the state of this line to guide the replacement decisions of other coresident lines of the region. 
ETR reduces the bandwidth overhead of state updates by 70\% and enables RRIP-AOB to achieve 18\% speedup, nearing Ideal RRIP-AOB performance.
These benefits are obtained with a storage overhead of less than 1KB SRAM. 

\vspace{.01in}

\begin{tcolorbox}[top=3pt,bottom=3pt]
{\bf Note:} A cache implementing ETR still fundamentally employs line-based replacement -- it simply opportunistically  exploits spatial locality when it exists to reduce state update costs. 
We compare with alternative page-based~\cite{unison} designs in Section~\ref{ssec:page}, and grouped-metadata~\cite{simcache} approaches in Section~\ref{ssec:line}.
\end{tcolorbox}

\vspace{-.02in}

Overall our paper makes the following contributions:

\vspace{.03in}

{\setlength{\parindent}{0cm} {\bf Contribution-1:} To our knowledge, this is the first paper to investigate intelligent replacement / bypass policies for direct-mapped DRAM caches. 
We propose a bypass version of RRIP~\textit{(RRIP-AOB)} suitable for caches with limited associativity.
However, we find an effective replacement policy for DRAM caches must optimize not only hit-rate but also state update cost. We introduce two properties, \textit{coresidency} and \textit{eviction-locality}, that can be exploited to reduce state update cost for implementing intelligent replacement.}

\vspace{.05in}

{\setlength{\parindent}{0cm} {\bf Contribution-2:}  We propose \textit{Efficient Tracking of Reuse (ETR)}, a design that performs updates for only a subset of lines and uses their state to guide the replacement decisions of other lines.  ETR reduces bandwidth overhead of updates by 70\%, improves speedup to 18\%, and requires only 512-bytes.}

\vspace{.05in}

{\setlength{\parindent}{0cm} {\bf Contribution-3:} We discuss how our concepts of RRIP-AOB and ETR can be applied to enhanced policies that rely on signature-information (SHiP~\cite{SHiP}) for further speedup.}

\vspace{.05in}

{\setlength{\parindent}{0cm} {\bf Contribution-4:} We show that RRIP-AOB and ETR are general techniques also applicable to set-associative implementations of DRAM caches.}

\ignore{
As memory systems scale, heterogeneous memory solutions consisting of a high-capacity memory (commodity DRAM or 3D XPoint~\cite{intel:3dxpoint}) and a high-bandwidth memory (3D-stacked DRAM~\cite{HBM}) will be critical to feed the ever-growing capacity and bandwidth demands of today's workloads. 
In fact, recent products such as Intel's Knights Landing (KNL)\cite{KNL} design already contain heterogeneous memory systems, consisting of a stacked DRAM based cache in front of commodity DRAM. In such systems, performance relies heavily on the DRAM cache hit-rate and bandwidth consumption -- improving DRAM cache performance will have meaningful impact on future systems with heterogeneous memories.
In this work, we investigate the potential of using intelligent cache replacement or bypass policies to improve DRAM cache hit-rate and bandwidth. Typical intelligent replacement policies rely on storing per-line state and updating state on cache events such as hits, misses, or bypasses. Such stateful policies are effective in on-chip caches with sufficient on-chip bandwidth to maintain state; however, such policies may not scale to off-chip DRAM caches as each state update would require an expensive off-chip DRAM update. This off-chip state update constitutes a significant bandwidth overhead that  traditional cache replacement / bypass policies typically have not optimized for. Thus, in DRAM Caches, the effectiveness of replacement or bypass policies are dictated not only by the hit-rate improvement, but also by the bandwidth consumption of the policy. Ideally, for DRAM caches, we desire a replacement / bypass policy that can achieve both intelligent replacement and minimal state update cost.




Commercial designs for DRAM cache organize the tag-store such that the tags can be kept in DRAM (to reduce storage overheads) and yet the tags can also be obtained with low-latency and low-bandwidth overheads~\cite{KNL}\cite{Alloy}. 
For example, KNL organizes the DRAM cache (we call this design {\em KNL-Cache}) as a direct-mapped structure and co-locates each tag with its data-line and streams out the tag and the data together in one access.
While such a limited-associativity design is effective for enabling low-latency and bandwidth-efficient tag access, a cache with limited associativity can suffer from significant conflict misses. We investigate the extent to which an intelligent replacement / bypass policy can reduce conflict misses for such commercial DRAM caches. 



To improve hit-rate for such limited-associativity caches, we can employ effective replacement policies that try to keep the most useful lines in the cache.
The most effective replacement policies are often re-use based~\cite{RRIP,SHIP,Hawk-eye,multiperspective} -- they exploit the observation that lines that have previously had reuse, will likely see future reuse and should be protected. 
We would like to use this class of replacement policies to improve hit-rate of commercial DRAM caches. However, two major problems preclude implementing such policies on today's DRAM caches. For one, reuse-based policies often need significant state update on hit, miss, or bypass that can limit their effectiveness on DRAM caches. Secondly, such replacement policies become ineffective when applied to limited associativity caches. For example, a direct-mapped cache would only have one victim and would degrade into Always-Install. And, a two-way cache would always need to find a victim, even if both lines are predicted to have high reuse. We can potentially avoid victimizing hot lines by intelligently utilizing \textit{Cache Bypassing}~\cite{BEAR,crc1,ctrbypass}. 
However, prior bypassing techniques on DRAM caches have been limited to global probabilistic-bypassing policies meant to protect the entire cache's working set (BAB~\cite{BEAR} is limited to 3\% speedup in Figure~\ref{fig:intro}(b)); whereas, we want a bypassing policy that can exploit the difference in characteristics of individual lines.

In this paper, we use {\em Re-Reference Interval Prediction (RRIP)}~\cite{RRIP} as a representative example of a replacement policy that is designed to exploit reuse~\cite{RRIP,SHIP,crc1,Hawk-eye,multiperspective}.  RRIP requires that each line is equipped with metadata bits (two-bit counter called RRPV) to track reuse. RRPV is set to 0 on a hit, and the victim line is identified as a line that has an RRPV of 3 (and if no lines have an RRPV of 3, all counters in the set are incremented and victim-selection is repeated). While RRIP is effective for set-associative caches, it becomes ill-defined for direct-mapped caches, as such a cache would have only one counter in the set. Following the algorithm for selecting a victim will always cause the resident line to get evicted, even if the line had an RRPV of 0. Similarly, bypassing the incoming line if the resident line has an RRPV=0 will mean that such lines will never get evicted from the cache,
even long after they have ceased to give any hits. 

To enable reuse-based replacement policies for direct-mapped DRAM cache, we propose a bypass version of RRIP, which we call \textit{RRIP-BYP}. The key mechanism in RRIP-BYP, is to age the counters on cache bypass. For example, if a good victim cannot be found (no RRPV are 3), we bypass the incoming line and age the reuse counters (all RRPV incremented). After several bypass+age events, a resident line that is no longer useful will have its RRPV reach 3 and become a candidate for eviction. This enables the cache to protect lines that have had reuse via bypassing, but also provides a path to eventually victimize cold lines. Our insight makes RRIP (and other reuse-based policies) applicable to DRAM caches.

}








\ignore{
GREY BOX: YES, lots of literature. Objective is not just byp policy, but reduce bw overhead associated with byp policy, as we need to track reuse.
OR: Prior work has used ctr-based bypassing. Requires updating state. State in DRAM costs significant bandwidth. Ineffective if high bandwidth cost. We additionally consider and deal with cost.
\footnote{Counter-based bypassing has been investigated previously\cite{ctrbypass}, but their counters must be updated on every cache event (e.g., hit) and constitute a significant bandwidth overhead under a DRAM cache. We additionally consider and reduce state update cost in our work.}
\footnote{Prior work has explored counter-based bypassing~\cite{ctrbypass}. But any intelligent replacement policy requires updating per-line state on hits, misses, or bypasses, which constitute a significant bandwidth overhead when applied to DRAM caches. Our work additionally accounts for and reduces state update cost.}
}



\ignore{
To the best of our knowledge, this is the first paper to investigate using intelligent reuse-based replacement policies for improving the hit-rate of practical direct-mapped DRAM caches. Notably, we show how to formulate RRIP as a bypassing policy with our proposed RRIP-DME.

However, RRIP needs significant per-line state, which would need impractically large amounts of SRAM storage overhead. We show we can get near to the ideal performance by storing state in DRAM and reducing bandwidth needed to maintain state in DRAM with our proposed ETR.

To enable reuse-based replacement policies for direct-mapped caches, this paper proposes Direct-Mapped Edition of RRIP (RRIP-DME) whereby we implement

}

\ignore{

\vspace{.05in}

Overall our paper makes the following contributions:

\vspace{.05in}

{\setlength{\parindent}{0cm} {\bf Contribution-1:} To the best of our knowledge, this is the first paper to investigate intelligent replacement policies for today's direct-mapped DRAM caches. We show how to design RRIP as a bypassing policy suitable for direct-mapped cache with \textit{RRIP: Direct-Mapped Edition (RRIP-DME)}. We also introduce two properties, \textit{coresidency} and \textit{eviction-locality}, that can be exploited to reduce state update cost for implementing intelligent replacement for DRAM caches.}

\vspace{.05in}

{\setlength{\parindent}{0cm} {\bf Contribution-2:}  We propose \textit{Efficient Tracking of Reuse (ETR)}, a design that keeps the replacement-state in DRAM and performs updates for only a subset of lines and uses their state to guide the replacement decisions of other lines.  ETR reduces bandwidth overhead of updates by 70\%, bridges 70\% of the gap between SIS and SID, and requires only 512 bytes.}

\vspace{.05in}

{\setlength{\parindent}{0cm} {\bf Contribution-3:} While we perform our studies using the RRIP replacement policy, we also discuss how our concepts of DME and ETR can be applied to enhanced policies that rely on signature-information (SHiP~\cite{SHIP}) for even higher performance improvement.}

\vspace{.05in}
{\setlength{\parindent}{0cm} {\bf Contribution-4:} While we perform our studies on a direct-mapped cache, we show that our concepts of DME and ETR can even be applied to associative cache design~\cite{ACCORD}.}

}





\ignore{
we find that as SHiP is an accurate predictor of reuse, we can s

to further reduce state update costs by avoiding state update when incoming line is predicted to have no reuse.

formulate SHiP as way to further reduce the bandwidth of state updates.

We show how to employ concepts of ETR on state-of-the-art signature-based replacement policies . 

In addition,

}

\ignore{
However, there are still cases where state update traffic is significant. If we can predict that the incoming line will not be reused, we can avoid sending demotions as the incoming line is unlikely to displace the resident line (due to no reuse). For example, prior proposals have used PC information\cite{SHIP} to predict reuse characteristics. Adding \textit{Signature-based reuse prediction (Sig-ETR)} enables further reduction in state update cost to finally improve 19.4\% speedup to 21\%.
}

\ignore{

}

\ignore{
\begin{figure}[htb] 
	\centering
	\includegraphics[width=1.00\columnwidth]{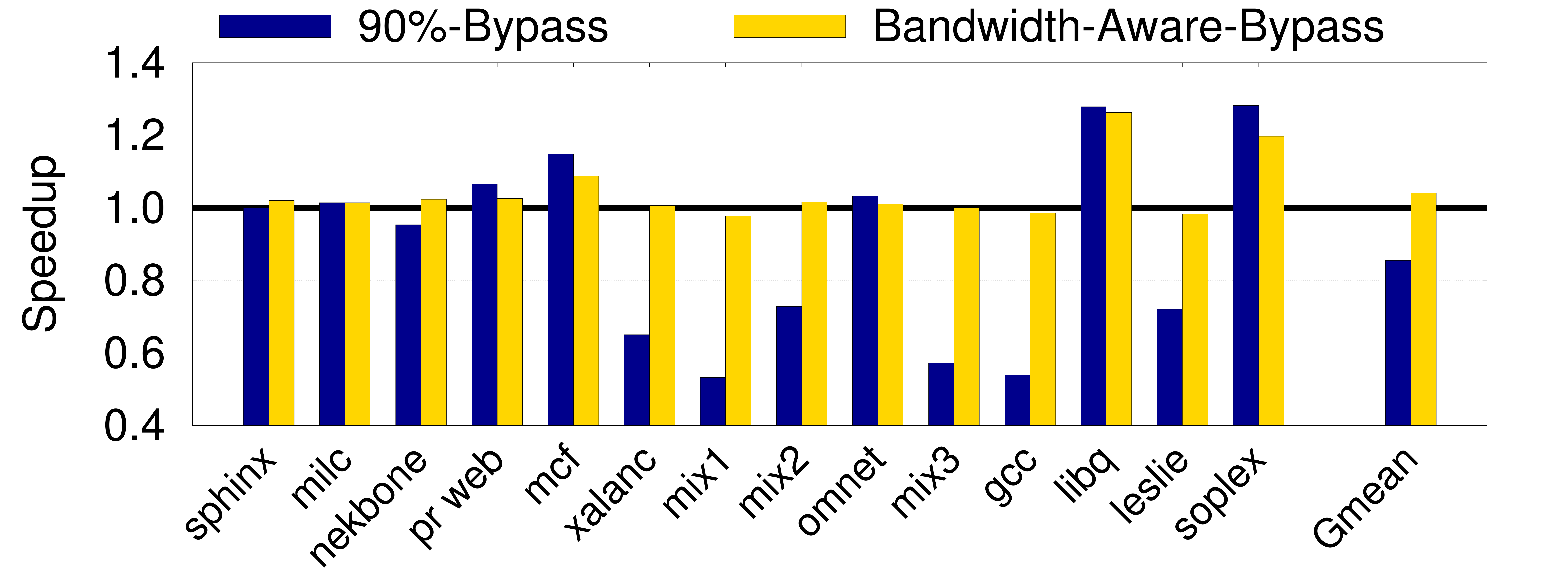}
	\vspace{-0.35in}
	\caption{Performance of DRAM cache employing 90\%-Bypass (16\% degradation) and Bandwidth-Aware Bypass (3\% speedup) on a system with NVM memory. Current bypassing solutions either sacrifice hit-rate for bandwidth or are oblivious to re-use of individual lines.}
	\label{fig:intro_number} 
	\vspace{-0.05in}
\end{figure}
}

\ignore{
	{\setlength{\parindent}{0cm} {\bf Contribution-4:} We then show how to provide Quality-of-Service guarantees under a shared DRAM cache. Through a combination of sampling and storing virtual tags in DRAM cache, we can observe hit-rate under a partitioned DRAM cache. We protect workloads that experience significantly lower hit-rate under shared DRAM cache to enable a QoS guarantee that no workloads experience  greater than 5\% degradation relative to partitioned cache.}
		\vspace{.05in}
}

\ignore{
Such a replacement policy purely optimizes for combined hit-rate and bandwidth for a \textit{shared} cache. However, if multiple workloads with different applications are running, some less-memory-intensive workloads may find it difficult to obtain space in the cache. In such mixed-application scenarios, we would like to obtain some level of Quality-of-Service(QoS) guarantees, where each core can achieve similar level of performance as if it had a \textit{partitioned} slice of the DRAM cache. To accomplish QoS, we estimate hit-rate under partitioned cache by sampling 1\% sets and virtually installing tags into DRAM-ECC space. We then compare estimated partitioned-cache hit-rate with current shared-cache hit-rate. If a core sees significantly degraded hit-rate under shared, we protect cache space for that core. By protecting degraded cores, we are able to ensure individual core performance to be at-worst similar to performance under a partitioned cache. By solving QoS, we are able to limit worst-case slowdown to within 5\%, and improve overall speedup across 30 mixed workloads.
}


\ignore{
A traditional method to improve cache hit-rate is to design an effective cache replacement policy. Cache replacement policies
can simultaneously improve both system-performance and system-power, at little cost. 
However, DRAM cache designs in the near future are likely to be direct-mapped to enable efficient tag-access. As such, prior work on cache replacement, which rely on finding good eviction candidates from multiple ways, can be difficult to apply on DRAM caches.

In addition, intelligent replacement policies often require remembering re-use characteristics (LRU\cite{lru2}, RRIP\cite{RRIP}, SHiP\cite{SHIP}). Maintaining re-use state in SRAM would need a large multi-megabyte storage (4-12MB). Meanwhile, maintaining re-use state in DRAM would need significant bandwidth to keep state updated (update on hits are significant bandwidth overhead). Replacement policies for DRAM cache will need to take into account both SRAM storage and DRAM bandwidth overheads.
}




\ignore{
In this work, we aim to gather the insights learned in prior last-level cache replacement strategies, and formulate them in a method suitable (storage and bandwidth-efficient) for direct-mapped DRAM caches by Efficiently Tracking Reuse (ETR). 
Figure~\ref{fig:intro}(c) shows prior bypassing solutions
are too coarse-grain (Bandwidth-Aware Bypass) to bypass the cache effectively\cite{BEAR}.
Instead, we formulate state-of-art RRIP and SHiP replacement policies as DRAM cache bypassing policies and reduce the state update bandwidth overheads, to improve hit-rate and reduce bandwidth consumption.
}




\ignore{

}


\ignore{
KNL design: explain how it works, problem of replacement, opportunity (extra bits available)
Replacement Policies:  Direct map can do bypassing, not enough. Recency is no good (LRU is Always Install). Need frequency or reuse based replacement.  RRIP
RRIP for Direct-Mapped: Explain how this works (upgrade/downgrade). CTR can be kept in SRAM or DRAM
Performance and Bandwidth: Show performance of PB, BAB, SRAM, DRAM and Bandwidth Breakdown of DRAM
Goal:  Performance of SRAM while keeping state in DRAM (share brief insight)
}

\begin{figure*}[htb]
		\centering
   \vspace{-0.15in}
	\centerline{\includegraphics[height= 1.4 in]{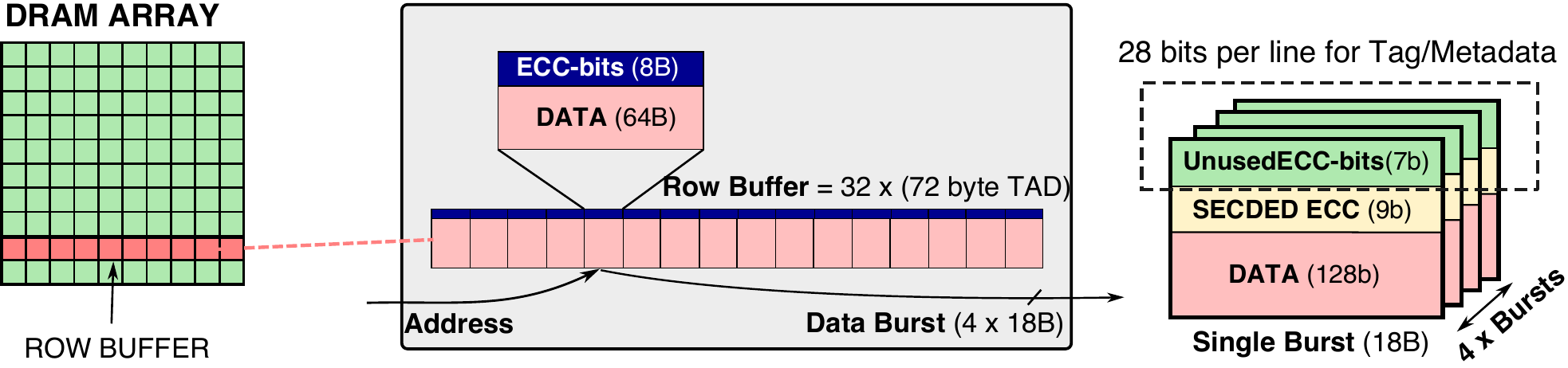}}
	\vspace{-0.05in}
	\caption{Organization of the DRAM cache used in KNL. DRAM cache is organized at a linesize of 64 bytes, is direct-mapped, and tags are kept with the data-line.  On an access, the DRAM cache transfers 72 bytes using four bursts on an 18-byte bus (16-bytes for data + 2-bytes for ECC). We need only 9-bits for SECDED on 16-bytes of data, which leaves 7 unused ECC bits in each burst that can be used to store metadata (KNL utilizes these 28 unused ECC bits to store tags).}
    \vspace{-0.1 in}
	\label{fig:alloy}
\end{figure*}

\newpage

\section{Background and Motivation}

We present the organization of our DRAM cache and discuss the storage and bandwidth
constraints that make it challenging to apply intelligent replacement policies.

\subsection{Organization of a DRAM Cache (KNL)}

As the tag storage required for gigascale DRAM caches is large, DRAM cache designs often store tags in DRAM and intelligently organize their structure to enable efficient tag-access. The baseline we use for this study is the direct-mapped, tags-in-ECC organization used in {\em Intel's Knights Landing (KNL)} design~\cite{KNL,Alloy}. Figure~\ref{fig:alloy} shows the organization of the DRAM cache in KNL. The DRAM cache places each tag information in the unused bits in the ECC space and streams out the data and tag (contained in ECC) on each access. The tag information is used to determine cache hit or miss. On a tag match, the data is available to service the request immediately, without any additional latency.  Thus, co-locating the tag and data allows the DRAM cache access to be serviced in just one DRAM request, which makes the cache hit operation both low-latency and bandwidth-efficient~\cite{Alloy}.  Our goal is to increase the hit-rate of such DRAM caches. In fact, the DRAM cache only uses about 8-10 bits from the unused 28 bits in the ECC space, so we have 18-20 bits per line available for managing the DRAM cache intelligently. We \ignore{seek to}leverage these bits to build intelligent replacement policies.


\subsection{Replacement / Bypass Policies for 1-Way}

Typically, cache replacement policies are discussed in the context of a set associative cache, as the set contains multiple lines and there is a choice of the line to evict.  For a direct-mapped cache, the set contains only one line,  so if we want to install, there is exactly one place the line can go, and we do not have a choice in selecting the victim. However, we could choose to bypass the line, so the binary choice for a direct-mapped cache becomes, whether to evict the resident line or to bypass the incoming line.  We can improve the hit-rate by making this binary decision intelligently. We explain different replacement strategies for a direct-mapped cache.

\vspace{0.05 in}
\noindent{\bf Probabilistic Replacement:} The simplest policy is to bypass the incoming line with a certain probability. For example, Bandwidth-Aware Bypass (BAB)~\cite{BEAR,crc1} bypasses the incoming line with 90\% probability to reduce install bandwidth, as long as hit-rate remains unaffected.  Figure~\ref{fig:bab} shows that such global bypassing policies are coarse-grain and miss out on bypassing opportunities that exploit per-line information.

\vspace{0.05 in}
\noindent{\bf Recency-Based Replacement:}  LRU\cite{lru2} installs incoming lines with the highest priority, based on the heuristic that recently-used lines are more likely to be re-used. On a direct-mapped cache, LRU degenerates into an {\em Always-Install} design, as the incoming line is the most recent. Enhancements of LRU, such as DIP~\cite{DIP}, degenerate to probabilistic bypass.

\vspace{0.05 in}
\noindent{\bf Reuse-Based Replacement:} Replacement policies that exploit reuse (also called re-reference or frequency) are resilient to thrashing and scans\cite{RRIP,freq1,VWAY}.  Such policies can protect the direct-mapped DRAM cache from thrashing when multiple pages are mapped to the same set of the DRAM cache.  We discuss {\em Re-Reference Interval Prediction (RRIP)}~\cite{RRIP} policy.

\begin{figure}[htb] 
	\centering
	\includegraphics[width=1.0\columnwidth]{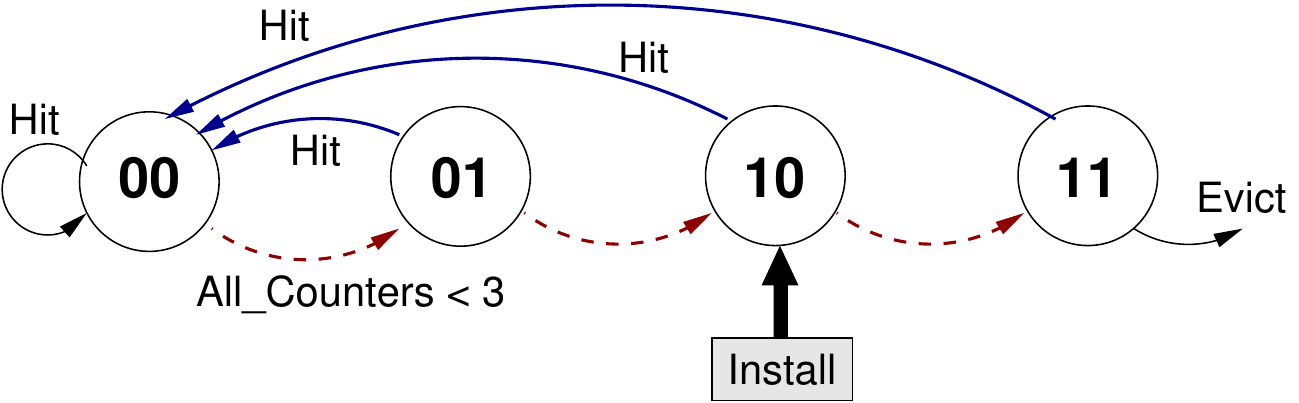}
	 \vspace{-0.15 in}
	 \caption{Re-Reference Interval Prediction (RRIP).} 
   \vspace{-0.05 in}
	\label{fig:rrip} 
\end{figure}

Re-Reference Interval Predictor\cite{RRIP} is a thrash-and-scan-resistant replacement policy often used in last-level caches. As shown in Figure~\ref{fig:rrip}, each line is equipped with a 2-bit counter to track the {\em Re-Reference Interval Prediction Value (RRPV)}.  On a hit to the line, the RRPV is \textit{Promoted} to 0. On a miss, the victim is found by searching from way 0 and finding the first line in the set with RRPV of 3. If no such line is found, the RRPV of all lines in the set is \textit{Demoted} (i.e., incremented) and the search is repeated. Lines are installed in RRPV=2 to protect the lines that were re-used.


\vspace{0.05 in}
\noindent\textbf{Challenge in Using RRIP for Direct-Mapped Cache:} Just like other replacement policies based on reuse-information, RRIP operates by comparing the counter values of multiple candidates in the set.  It becomes ill-defined for a direct-mapped cache, where there is only one counter, which means the resident line will always get evicted regardless of the past behavior.  Thus, for a direct-mapped cache RRIP degenerates into always-install (or always-bypass if the incoming line is bypassed unless the RRPV of the resident line equals 3). We propose extensions that make reuse-based policies viable for direct-mapped and two-way caches, and implementations that reduce the cost of tracking the RRPV state for gigascale DRAM caches. We discuss our solution after methodology.

\ignore{

We can implement RRIP by storing Replacement-State in DRAM (RRIP-RSD). 
For example, we can store the 2-bit RRPV of RRIP alongside Tag in the ECC space. However, such a policy requires significant bandwidth to \textit{promote} state on hit, and \textit{demote} state on bypass. 
We can alternately try to implement RRIP by storing Replacement-State in SRAM (RSS-RRIP). However, storing just 2-bits per every cache line requires 8MB of storage. There is currently no policy that can achieve all of our goals simultaneously.

\begin{table}[htb]
    \vspace{-.2in}
	\centering
	\caption{Comparison of Replacement Policies}
	\label{tab:repl}
	\renewcommand{\arraystretch}{.80}
		\setlength{\tabcolsep}{4.5pt}{
	\setlength{\extrarowheight}{2pt}{
	\begin{tabular}{|c||c|c|c|c|}
		\hline
		\multirow{2}{*}{}& \multirow{2}{*}{Miss-rate} & \multirow{2}{*}{Install-BW} & {SRAM}            & {DRAM} \\ 
        &&&Storage-Cost&Update-Cost \\ \hline \hline
		Always-Install & Med     & High       & None                    & None        \\ \hline
		Bypass-90\%          & High      & Low        & None                    & None        \\ \hline
		BAB              & Med     & Med        & Low       & None        \\ \hline
		RRIP-RSD       & Low     & Low        & Low       & High        \\ \hline
        RRIP-RSS       & Low     & Low        & High  & Low         \\ \hline
		Goal                 & Low     & Low        & Low                     & Low         \\ \hline
	\end{tabular}
}
}
\end{table}


\section{Tuning Replacement Policies for Bypassing}

We analyze insights behind prior works on replacement policies, and formulate bypassing versions of them for DRAM cache to evaluate their potential. Our baseline is Always-Install incoming lines.

\subsection{Bimodal Insertion Policy (BIP)}

The main idea of Bimodal Insertion Policy (BIP)~\cite{DIP} is to protect working set against thrashing when the number of ways needed is larger than the number of ways available. It does this primarily by installing most lines at lowest priority and filtering them out quickly. For example, if there are 8-ways, 7-ways will be protected and kept, while the 8-th way will be kept for low-priority insertions and will have constant evictions. This maintains a hit-rate of roughly 7 / N-ways-needed, when N-ways-needed are larger than 7. We implement a simple version of BIP by simply bypassing (i.e., not installing into) the DRAM cache 90\% of the time. This affords working-set protection capabilities, while allowing for some small chance of learning new working sets. However, the bypassing version of the policy still suffers from learning new working sets slowly.

\setcounter{figure}{1}
\begin{figure*}[htb] 
	\centering
	\includegraphics[height=1.5 in]{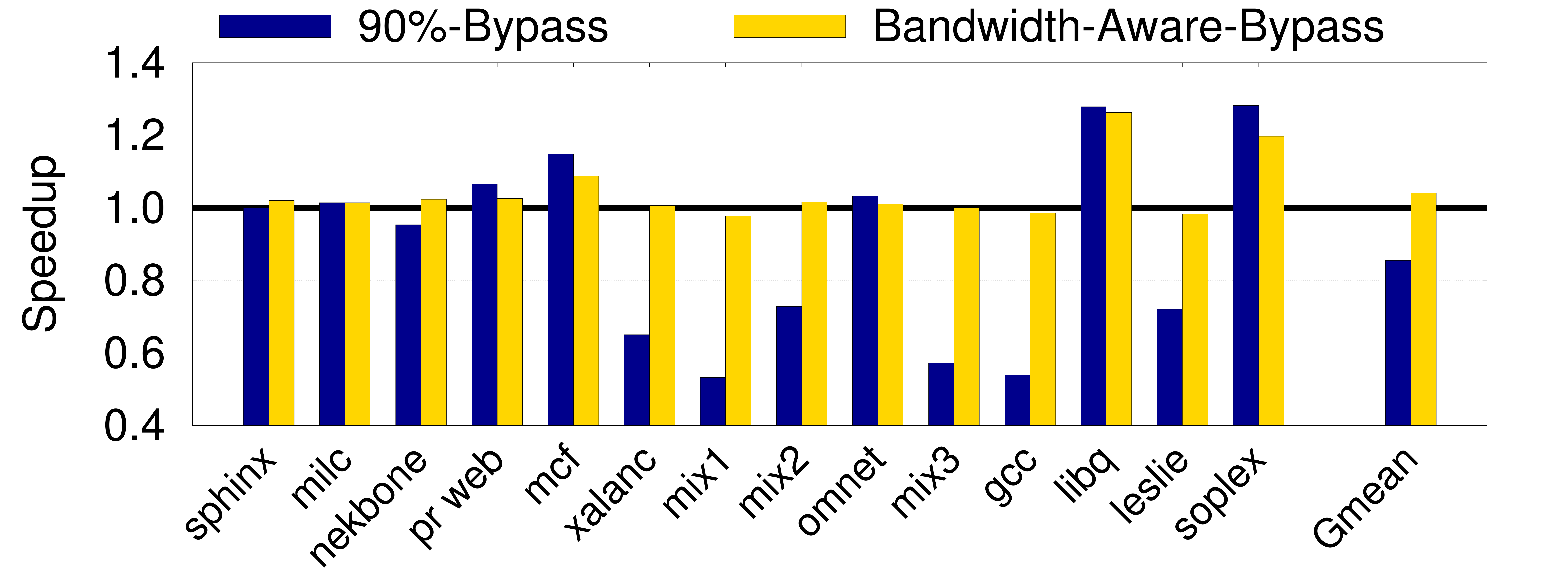}
	
	\caption{Performance of bypassing versions of BIP, DIP, TADIP, and SHiP++. SHiP++ improves performance by 9.8\%.}
	\label{fig:bypass_perf} 
\end{figure*}

\subsection{Dynamic Insertion Policy (DIP)}

The main of idea of Dynamic Insertion Policy (DIP)~\cite{DIP} is to switch between learning quickly (baseline insertion policy) and protecting working set (BIP) to provide better performance than either. DIP estimates which policy is more effective by sampling a small number of sets to use either policy, and determining which policy provides higher hit-rate by having them increment and decrement a counter based on the number of misses each policy causes. The more effective policy will see the counter saturated in the other direction, and the entire cache will switch to using the more effective policy. We apply a similar technique by sampling 1/64 (1.5\%) of the lines to estimate effectiveness of Always-Install vs. BIP and globally enforce the better-performing policy. DIP achieves better than either bypass-BIP or Always-Install. We evaluate a thread-aware version of DIP called TADIP~\cite{TADIP} as well.

}

\ignore{
	Replacement Policies
	BIP. DIP. TADIP. Protect working set. Useful. Can apply for bypassing as well.
	But SHiP can be. SHiP protects and bypasses lines based on signature (i.e. PC). We show a bypassing version of this as well.
}

\ignore{
Unfortunately, it is challenging to make such practical DRAM cache design set-associative, as the tags are present with each line, so making the cache two-way set associative would mean streaming  would need to stream two lines on each access to make the cache two-way set-associative.
}

\newpage
\section{Methodology}
\vspace{.1in}
\subsection{Framework and Configuration}
\label{subsection:conf}

We use USIMM~\cite{USIMM}, an x86 simulator with detailed memory
system model.  We extend USIMM to include a DRAM cache.
Table~\ref{table:config} shows the configuration used in our study.
We model a configuration similar to a Intel Knights Landing (KNL) Sub-NUMA Cluster (one-eighth size).
We assume a four-level cache hierarchy (L1, L2, L3 being on-chip SRAM
caches and L4 being off-chip DRAM cache). All caches use 64B line
size. We model a virtual memory system to perform virtual to physical
address translations.  The L4 is a 2GB DRAM cache~\cite{sodani_2015,Alloy}, which is direct-mapped and places tags with data in the unused ECC bits. The parameters of our DRAM
cache is based on HBM technology~\cite{HBM}.
The main memory is based on non-volatile memory
and assumed a latency similar
to PCM and 3D-XPoint\cite{xpoint_latency2,optane_latency,intel:3dxpoint,pcmbook,PCM:ISSCC2012,Wong:pcm,lee:isca09}:
the read latency is 4X  
that of DRAM~\cite{ddr4}, and write bandwidth is worse than read bandwidth. We perform evaluations with DRAM-based memory in Section~\ref{ssec:dram-memory}.

\begin {table}[!h]
\vspace{-.17in}
  \begin{center}
      \caption{System Config (KNL \sfrac{1}{8} Sub-NUMA Cluster)}
          \vspace{0.09 in}
\resizebox{3.225in}{!}{
\renewcommand{\arraystretch}{.80}
\setlength{\extrarowheight}{2pt}{

      \begin{tabular}{|l|l|}  \hline 

Processors & 8 cores; 3.0GHz, 2-wide OoO \\
Last-Level Cache & 8MB, 16-way \\ \hline

\multicolumn{2}{|c|}{\bf DRAM Cache } \\ \hline

Capacity               &    2GB         \\
Bus Frequency          &     500MHz (DDR 1GHz) \\
Configuration       &  4 channel, 128-bit bus \\
Aggregate Bandwidth              &      64 GB/s \\ 
tCAS-tRCD-tRP-tRAS     &      13-13-13-30 ns \\ \hline

\multicolumn{2}{|c|}{\bf Main Memory (PCM) } \\ \hline

Capacity               &    64GB        \\
Bus Frequency          &     1000MHz (DDR 2GHz) \\
Configuration       &  1 channel, 64-bit bus \\
Aggregate Bandwidth              &      16 GB/s \\ 
tCAS-tRCD-tRP    &      13-128-8 ns \\
tRAS-tWR     &          143-160 ns \\ \hline

      \end{tabular}
}
}
      \label{table:config}
\vspace{-0.15 in}

      \end{center}
    \vspace{-0.05 in}
\end{table}


\subsection{Workloads}
\label{subsection:workloads}

We use a representative slice of 2-billion instructions selected by
PinPoints~\cite{pinpoint}, from benchmarks suites that include SPEC
2006~\cite{SPEC2006}, GAP~\cite{GAP}, and HPC.  
For SPEC, we pick a sample of high intensity workloads that have at least two miss per thousand instructions (MPKI).
The evaluations execute benchmarks in rate mode, where all
eight cores execute the same benchmark. In addition to rate-mode
workloads, we also evaluate 24 mixed workloads, which are created
by randomly choosing 8 of the 15 SPEC workloads that have at least two
MPKI. Table~\ref{table:benchmarks}
shows L3 miss rates, and memory footprints for the 8-core rate-mode workloads in our study. 

We perform timing simulation until each benchmark in a workload
executes at least 2 billion instructions.  We use weighted speedup to measure aggregate performance of the workload normalized to the baseline and report geometric mean for the average speedup across all the 21 workloads (11 SPEC, 4 SPEC-mix, 5 GAP, 1 HPC). Note that to keep the graphs readable we only use 4 mixed workloads for all of our results. However, we provide key performance results for the set of the remaining  20 mixes in Section~\ref{ssec:mixed}.

\begin {table}[!h]
\caption{Workload Characteristics}
\begin{center}
\vspace{0.05 in}
\resizebox{2.575in}{!}{
\renewcommand{\arraystretch}{.75}
\setlength{\extrarowheight}{2.4pt}{
        \begin{tabular}{|c|c|c|c|} \hline

Suite & Workload & L3 MPKI & Footprint \\ \hline \hline

\multirow{11}{*}{SPEC} & soplex & 35.3 & 1.8 GB \\ \cline{2-4}
& leslie & 22.1 & 623 MB \\ \cline{2-4}
& libq & 30.1 & 256 MB \\ \cline{2-4}
& gcc & 108.5 & 1.5 GB \\ \cline{2-4}
& omnet & 29.1 & 1.2 GB  \\ \cline{2-4}
& wrf & 10.4 & 1.1 GB \\ \cline{2-4}
& zeus & 7.0 & 1.6 GB \\ \cline{2-4}
& xalanc & 7.4 & 1.5 GB  \\ \cline{2-4}
& mcf & 101.1 & 13 GB  \\ \cline{2-4}
& milc & 31.2 & 4.5 GB  \\ \cline{2-4}
& sphinx & 15.0 & 146 MB  \\ \hline
\multirow{6}{*}{GAP} & cc twitter & 116.8 & 9.3 GB  \\ \cline{2-4}
& bc twitter & 101.2 & 13.5 GB \\ \cline{2-4}
& pr twitter & 126.6 & 15.3 GB    \\ \cline{2-4}
& pr web & 24.8 & 15.1 GB  \\ \cline{2-4}
& cc web & 11.4 & 9.3 GB  \\ \hline
HPC & nekbone & 13.71 & 44 MB  \\ \hline

        \end{tabular}
}
}
\vspace{-0.1 in}
      \label{table:benchmarks}
    \end{center}
\end{table}


\newpage

\section{RRIP: Age-On-Bypass}


If we want to use RRIP on direct-mapped DRAM caches, we have to solve two issues: how do we formulate RRIP as a bypassing policy suitable for caches with limited associativity, and how can we mitigate the state update cost of maintaining per-line reuse state in DRAM.


\subsection{RRIP as a Bypassing Policy}

\begin {figure}[!h]
	\centering
	\includegraphics[width=.95\columnwidth]{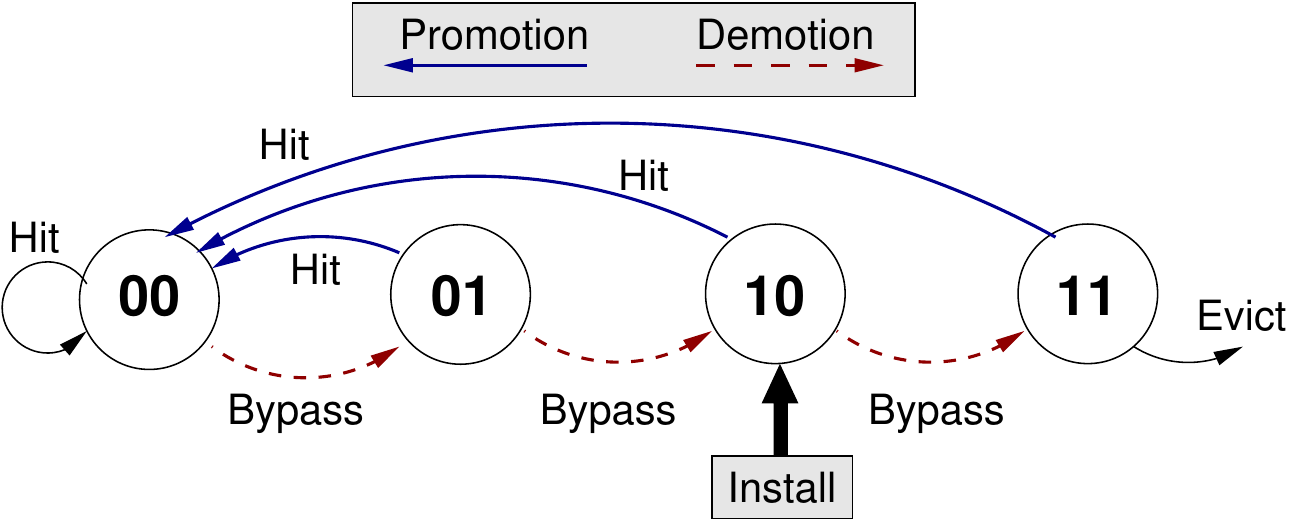}
	 \vspace{-0.10 in}
	 \caption{Overview of RRIP: Age-On-Bypass (RRIP-AOB). The transition from one state to another is accomplished with replacement-state update operation. Such updates may consume significant bandwidth.}
	\label{fig:rrip-dme} 
\end{figure}

We design a version of RRIP for limited-associativity caches, called \textit{RRIP: Age-On-Bypass (RRIP-AOB)}.  The key insight in RRIP-AOB is to use the episode of cache bypassing to age / update the RRPV information associated with the line. Figure~\ref{fig:rrip-dme} shows the overview of our design. RRIP-AOB needs to similarly track lines that have reuse, so RRIP-AOB Promotes state (sets RRPV to 0) on hit. RRIP-AOB can protect these reused lines by bypassing when reuse has been seen (bypass when RRPV is 0, 1, or 2). However, reused lines can now stay stuck in high priority state. We need a different mechanism to age older lines so that new lines can eventually be installed. We choose to implement aging by Demoting (increment RRPV) state when an incoming line is bypassed. This allows lines to naturally age to RRPV of 3, and be evicted in favor of the incoming line. Similar to RRIP, RRIP-AOB needs 2 bits per line to track RRPV.  A practical design must address where to store the RRPV bits and address the bandwidth needed to track the per-line RRPV.







\begin {figure*}[!ht]
    \vspace{-.05in}
	\centering
	\includegraphics[width=.94\linewidth]{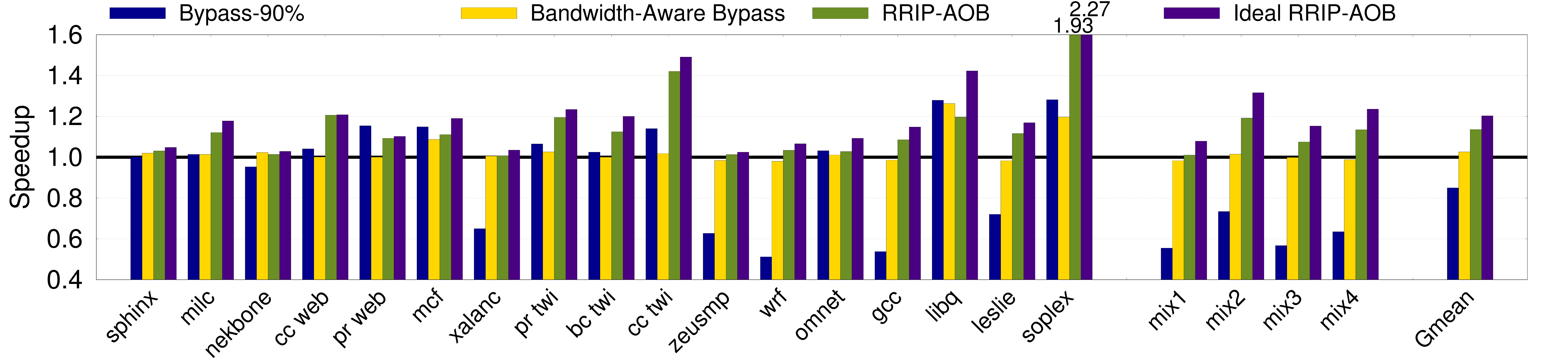}
   \vspace{-.1in}
    \caption{Speedup from different replacement policies over the baseline always-install direct-mapped DRAM cache. (a)~Bypass-90\% causes 15\% degradation, (b) Bandwidth-Aware Bypass provides 3\% speedup, (c)$\,$RRIP-AOB that maintains state in DRAM provides 13\% speedup, and (d) Ideal RRIP-AOB with no state update cost provides 20\% speedup}

	\label{fig:bab} 
    \vspace{-.1in}

\end{figure*}
\newpage

\subsection{Storing RRPV in DRAM}

A straight-forward way of incorporating RRIP into a DRAM cache is to extend the tag-entry of the line to incorporate the RRPV bits. We refer to this design as simply {\em RRIP-AOB}.
However, such a design incurs bandwidth overhead for performing update of the replacement state. Note that these accesses for updating the replacement state are not present in the baseline and for designs that do bypassing without tracking per-line state. 


Alternatively, we can avoid the bandwidth of replacement updates by storing the replacement state in a dedicated SRAM array. Unfortunately, for our 2GB DRAM cache, maintaining  2 bits of RRPV per line would need 8MB of SRAM, which is impractically large. We call this design \textit{Ideal RRIP-AOB}. 


\begin{figure}[htb] 
\vspace{-0.05 in}
	\centering
	\includegraphics[width=1.0\columnwidth]{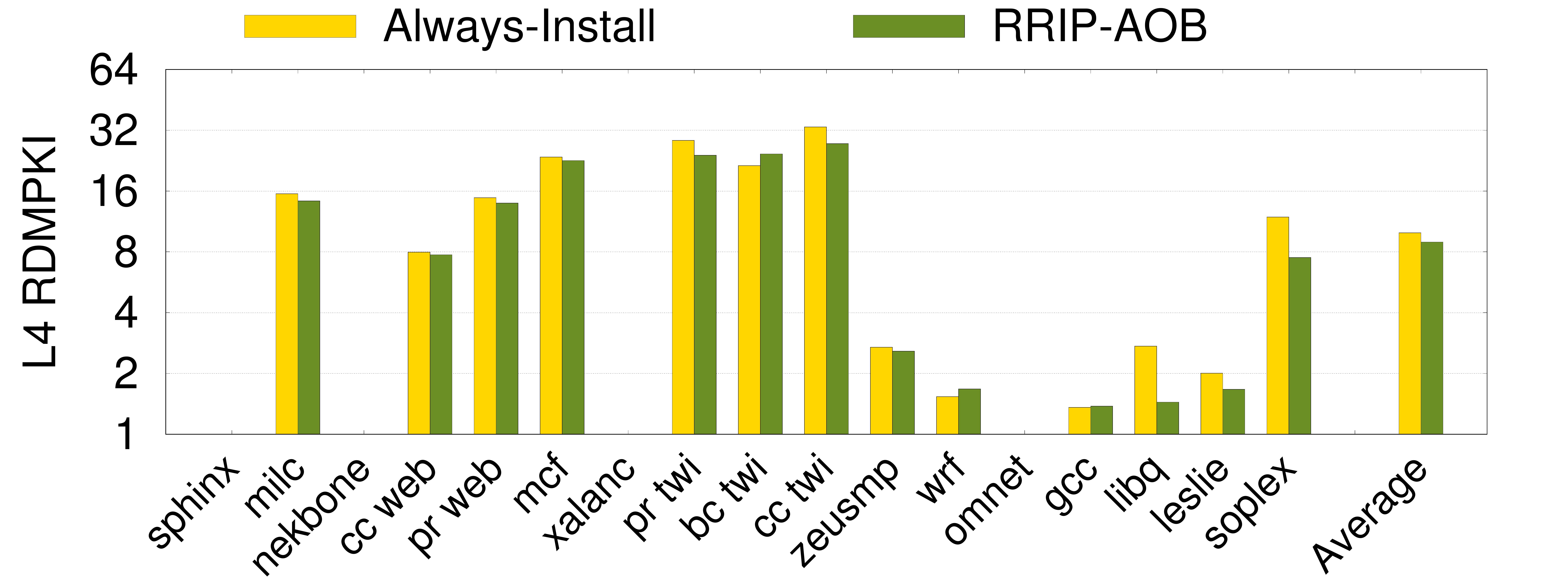}
    \vspace{-0.2 in}
	\caption{MPKI of baseline DRAM cache and RRIP-AOB. RRIP-AOB reduces misses by 10\%.}
	\label{fig:rdmpki} 
   \vspace{-0.15 in}
\end{figure}

\subsection{Benefits from Reuse-Based Replacement}

Intelligent replacement policies improve performance by reducing cache misses.  Figure~\ref{fig:rdmpki} shows the Misses Per Thousand Instructions (MPKI) for our baseline DRAM cache and with RRIP-AOB. RRIP-AOB reduces 10\% of the misses on average. However, the speedup from RRIP-AOB also depends on bandwidth used in replacement-state updates.

Figure~\ref{fig:bab} shows the speedup from different bypassing policies implemented on our 2GB DRAM cache. Performance numbers are normalized to the always-install policy.  Indiscriminately bypassing 90\% of the lines (Bypass-90\%) causes a degradation of 15\%. The adaptiveness of Bandwidth-Aware-Bypass (BAB)~\cite{BEAR,crc1} avoids slowdowns; however, the average speedup is only 3\%.  
With RRIP-AOB, the performance benefits is 13\%, whereas with Ideal RRIP-AOB the speedup could be 20\%. 
Thus, there is significant room for performance improvement with reuse-based replacement policies. Unfortunately, obtaining this benefit in a practical manner is challenging as maintaining accurate per-line state in DRAM requires significant bandwidth for state updates.


\subsection{Dissecting BW of Replacement-Updates}

\begin{figure}[htb] 
	\centering
    \vspace{-0.20 in}
    \includegraphics[height = 1.7in]{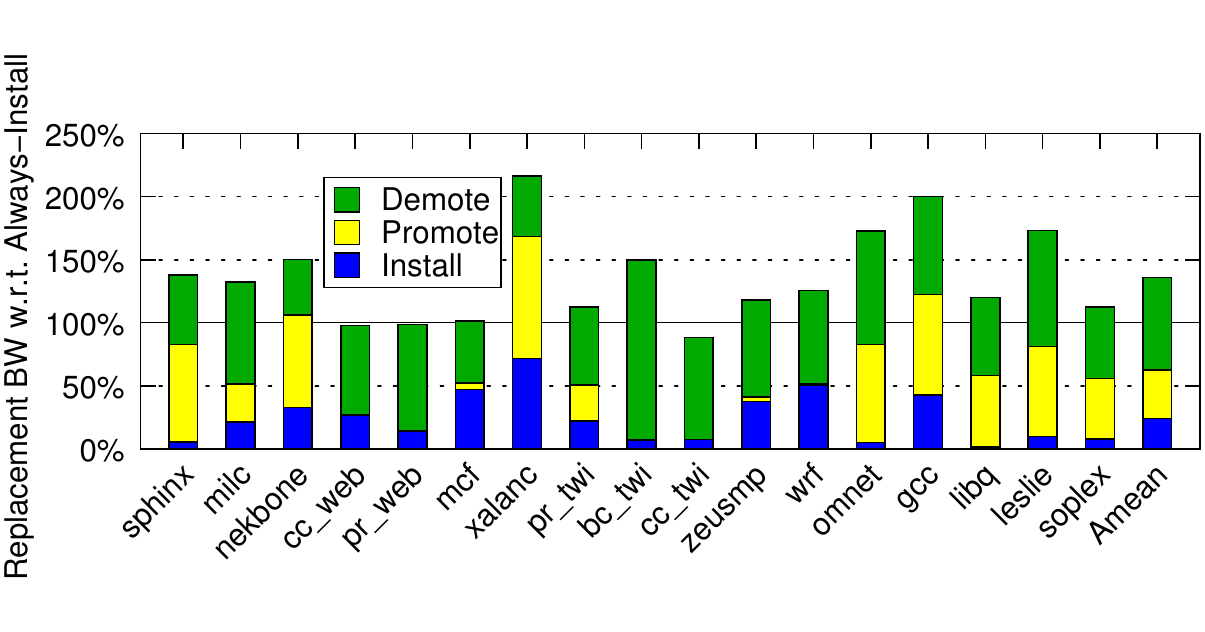}
    \vspace{-0.22 in}
	\caption{Replacement bandwidth (Install, Promote, Demote) of RRIP-AOB, normalized to replacement bandwidth (Install) of Always-Install. RRIP-AOB reduces install bandwidth but incurs state update bandwidth.}
     \vspace{-0.14 in}
	\label{fig:rrip_bw} 
\end{figure}

To highlight the bandwidth differences between Always-Install and RRIP-AOB, we show the bandwidth needed to implement replacement policy for Always-Install and RRIP-AOB. Always-Install simply has install bandwidth, whereas RRIP-AOB additionally needs bandwidth to \textit{promote} and \textit{demote} state.
Figure~\ref{fig:rrip_bw} shows the replacement bandwidth
of RRIP-AOB, normalized to the replacement bandwidth of Always-Install. Of particular note, \textit{RRIP-AOB has the potential to save 76\% of the install bandwidth} (due to bypass), which can improve performance. However, it has overall increased bandwidth consumption due to \textit{promotion} and \textit{demotion}. If we want to obtain most of the benefits of RRIP, we must develop methods to reduce this bandwidth overhead.


\subsection{Potential for Improvement}

RRIP-AOB with state in DRAM is a practical design as it does not require any SRAM overheads, and can be implemented without any changes to the DRAM cache (the extra bits for RRPV are taken from the unused ECC bits).  However, it has two-thirds the speedup compared to the potential benefit of Ideal RRIP-AOB with no state update costs.  
Ideally, we would like to get speedup similar to Ideal RRIP-AOB, while needing low SRAM cost similar to RRIP-AOB. The goal of the next section is to develop such a solution.


RRIP-AOB simply suffers from high DRAM state update cost. If we can find effective ways to mitigate this bandwidth overhead, we can get most of the benefits at little cost. We develop an insight that if we can do replacement updates in an efficient manner for only a subset of the lines, then we can reduce the bandwidth for replacement updates and still retain most of the benefits. 


\section{Efficient Tracking of Reuse}

Demoting state on every cache bypass incurs significant bandwidth overheads--even if we choose to bypass the line, we still have to spend bandwidth to \textit{demote} the replacement-state. We can avoid state update costs if we have an effective way to infer an RRPV state. Our design reduces the  bandwidth consumed in performing updates of the replacement state by doing the updates for only a subset of the lines and using their replacement state to infer the replacement state of the other lines.  Our solution is based on two key properties,  {\em Coresidency} and {\em Eviction-Locality}, which we describe next.








\subsection{Insight: Coresidency and Eviction-Locality }

\begin{figure}[htb] 
	\vspace{-0.14in}
	\centering
	\includegraphics[width=1.0\columnwidth]{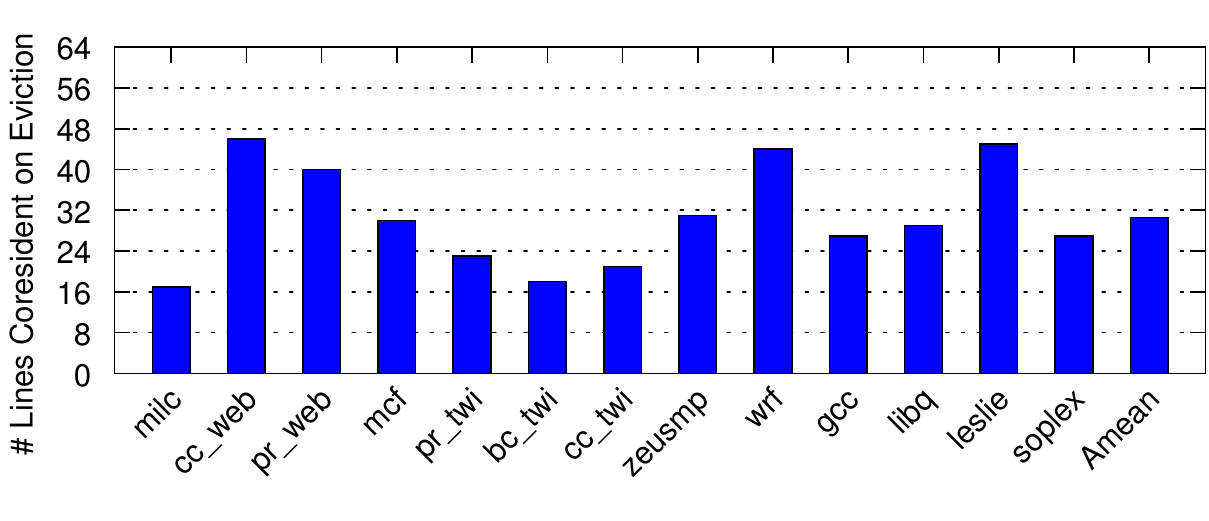}
	\vspace{-0.38in}
	\caption{Coresidency in DRAM caches.  Average number of coresident lines in a 4KB region on first line evicted from a region (workloads with L4 MPKI\textgreater1).}
	\label{fig:coresident} 
	\vspace{-0.1in}
\end{figure}

{\em Coresidency} indicates that at any given time if a line is present, then several other lines belonging to that region are also present in the cache. 
Coresidency indicates that there is some amount of spatial locality in the reference stream, even if such spatial locality is not perfect. A 4KB region contains 64 lines each of 64 bytes. Therefore, the maximum number of coresident lines for a region would be 63. Figure~\ref{fig:coresident} shows the level of coresidency for our workloads. In general, the workloads have between 16 to 45 coresident lines. We note that although this is lower than perfect spatial locality, there are still a large number of lines coresident (even 4 coresident lines can amortize 75\% state update cost). This shows there is potential for using one line to infer replacement state of many coresident lines.




\begin{figure}[bp] 
	\vspace{-0.17in}
	\centering
	\includegraphics[width=1.0\columnwidth]{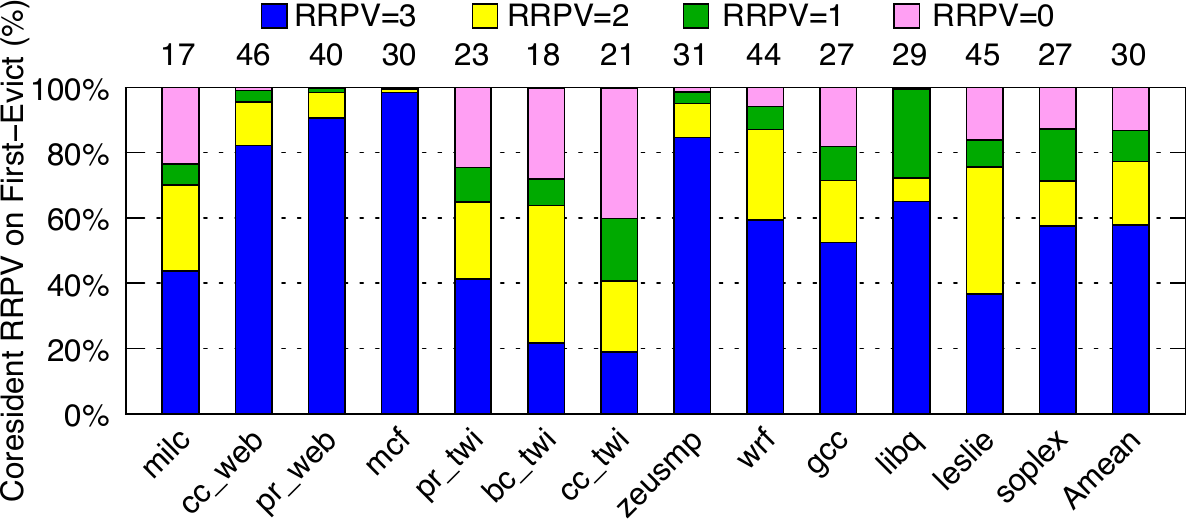}
	\vspace{-0.28in}
	\caption{Distribution of RRPV of coresident lines on first line evicted from a 4KB region, for workloads with L4 MPKI\textgreater1. Average number of coresident lines shown above workloads. Eviction of one line indicates other lines in a region are likely to be evicted soon (RRPV$\geq$2).}
	\label{fig:mpki_r} 
	\vspace{-0.1in}
\end{figure}

{\em Eviction-Locality} indicates that when a line gets evicted from the cache, then the replacement-state of the other coresident lines belonging to that region tend to have similar replacement-state as the line being evicted. Figure~\ref{fig:mpki_r} shows the distribution of the RRPV of coresident lines, on an eviction from L4 (on the first line evicted from a region). 


\vspace{-.05in}
\begin{tcolorbox}[top=1pt,bottom=1pt,left=2pt,right=2pt]
\noindent  \textbf{Combined insight:} On eviction, we typically observe 30 or more lines are coresident.
In addition, we find the coresident lines generally have similar RRPV state (77\% have RRPV$\geq$2).
Together, this means that if we maintain accurate RRPV for just one of the coresident lines, then we can infer RRPV state for the rest of the coresident lines in the region with reasonable accuracy. Our solution is based on exploiting this insight.
\end{tcolorbox}
\vspace{-.1in}



\subsection{Insight: Update Only the Representative}

We propose \textit{Efficient Tracking of Reuse (ETR)} to reduce bandwidth overheads of doing replacement updates in DRAM. We implement ETR on top of RRIP-AOB as an example.  ETR exploits the properties of coresidency and eviction-locality.  Instead of updating replacement state for all the lines in a region, ETR updates the state of only one {\em Representative-Line} among all the coresident lines.  The state of the Representative-Line is then used to guide the replacement policy of the coresident lines. The design of ETR consists of three parts: (1)  Selecting a Representative-Line in the region (2) Keeping accurate RRPV for \textit{only} the Representative-Line, and (3) Using the representative's RRPV to infer coresident lines' RRPV to make bypass decisions.  

\begin{figure}[htb] 
	\vspace{-0.11in}
	\centering
	\includegraphics[width=\columnwidth]{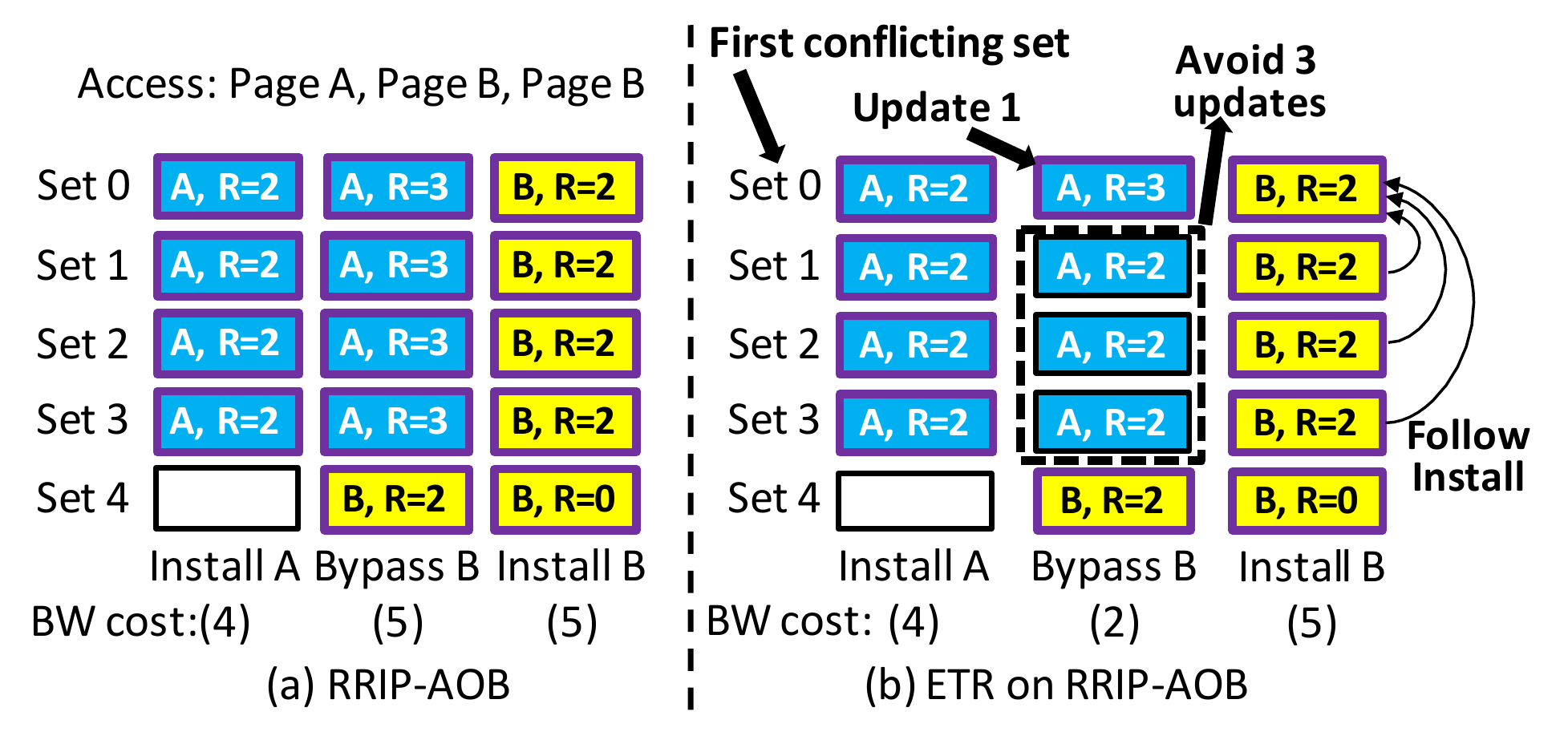}
	\vspace{-0.3in}
	\caption{ETR's representative-update and bypass-decision following enables similar RRIP-AOB install policy, at reduced update bandwidth (dashed box = benefit). }
	\label{fig:rrip_spatial_why} 
	\vspace{-0.09in}
\end{figure}

To implement representative-update, we first need to pick a stable representative line. Prior work finds the first access to a region is relatively consistent~\cite{sms}. If we maintain state for just the first conflicting set in a region, we can maintain good reuse information for the rest of the region without incurring extra bandwidth costs. Figure~\ref{fig:rrip_spatial_why} shows an example of how ETR's RRPV-inference (i.e., representative-update and bypass-decision following) can be used to obtain similar install-policy and hit-rate at reduced update cost. 

If we first access 4 lines from region A at time 0, we install region A with RRPV=2. If 5 lines from region B are then accessed, Figure~\ref{fig:rrip_spatial_why}(b) shows that we can save bandwidth and \textit{demote} only the state of the first conflicting set (being set 0). On second access to region B, set 0 with its RRPV of 3 will inform us that that region A was not used recently. This means that lines corresponding to region A have low reuse and should be evicted in favor of installing region B. We can then follow the region B install-decision for the rest of the lines. Such a policy will end up installing all of region B and result in an install policy similar to if we had maintained each state individually in Figure~\ref{fig:rrip_spatial_why}(a). As such, we can keep similar install policy and save update bandwidth with representative state update and bypass-decision following.





\begin{figure*}[htb] 
	\centering
	\includegraphics[height=1.5in]{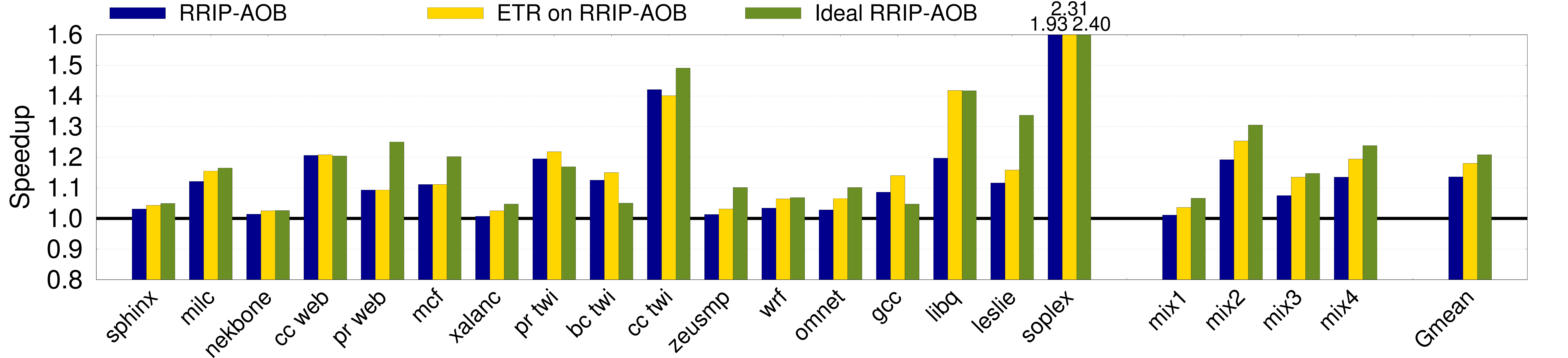}
	\vspace{-0.1in}
	\caption{Performance of RRIP-AOB, ETR on RRIP-AOB, and an Ideal RRIP-AOB with no state update costs. Coordinating bypass decisions with ETR reduces state update needs, and enables RRIP-AOB to obtain 18\% speedup.}
	\label{fig:rrip_perf} 
	\vspace{-0.15in}
\end{figure*}





\begin{figure}[htb] 
	\vspace{-0.1in}
	\centering
	\includegraphics[height=1.5in]{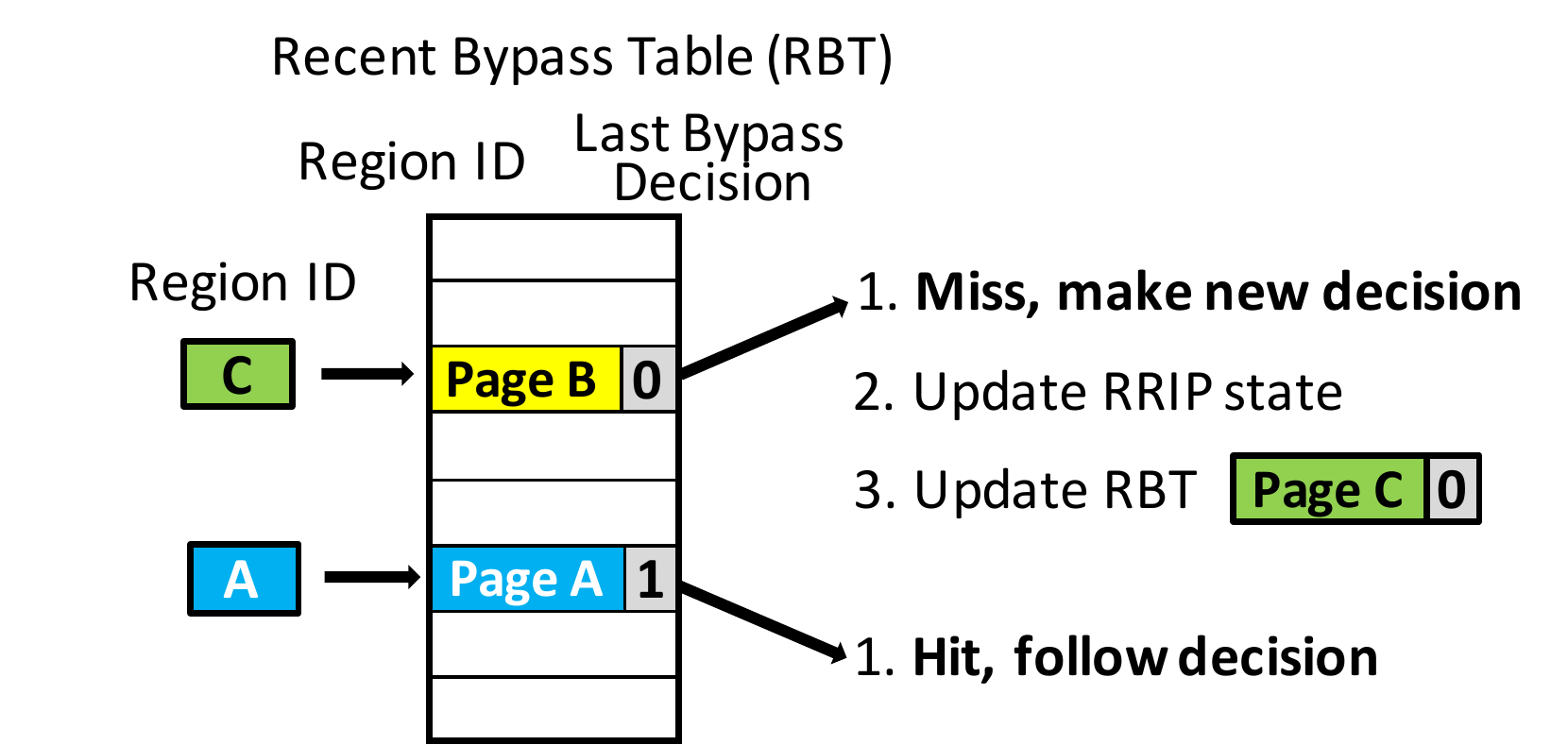}
	\vspace{-0.15in}
	\caption{Design of Recent-Bypass-Table to enforce coordinated-bypass and coordinated-state-update. Demotions only occur on first miss to a region.}
	\label{fig:rrip_spatial} 
	\vspace{-0.15in}
\end{figure}

\noindent\textbf{Structures for ETR:} To implement representative state update and bypass-decision following, ETR maintains a {\em Recent-Bypass Table (RBT)}, in Figure~\ref{fig:rrip_spatial}. RBT tracks recently seen regions ({\em Region-ID}) and the bypass decision made for them ({\em Last-Bypass-Decision}). RBT enables us to find the representative first-conflicting-set in a region (as the first conflicting set would have miss in RBT), keep just that set's RRPV up-to-date, and remember the first-conflicting-set's bypass decision to inform bypass decision for the other lines in the region (as the follower sets would hit in RBT and see previous decision made). We use a 128-entry RBT, which requires \textless 512B of SRAM (performance is relatively insensitive to RBT sizing).



\vspace{0.05 in}
\noindent\textbf{Operation of ETR:} On cache miss, we index into RBT with Region-ID. If there is an RBT miss, we are currently accessing the representative first-conflicting-set in a region. In this case, we should make a bypass decision based on its RRPV, spend bandwidth to demote state if bypass was chosen, and update the RBT so later accesses can make an informed bypass decision. Otherwise, if there is an RBT hit, the region has been recently accessed and already had a bypass decision made, so we should follow the {\em Last Bypass Decision} to keep similar install policy and save on demotion bandwidth.



\begin{figure}[htb] 
	\vspace{-0.18in}
	\centering
	\includegraphics[width=1.0\columnwidth]{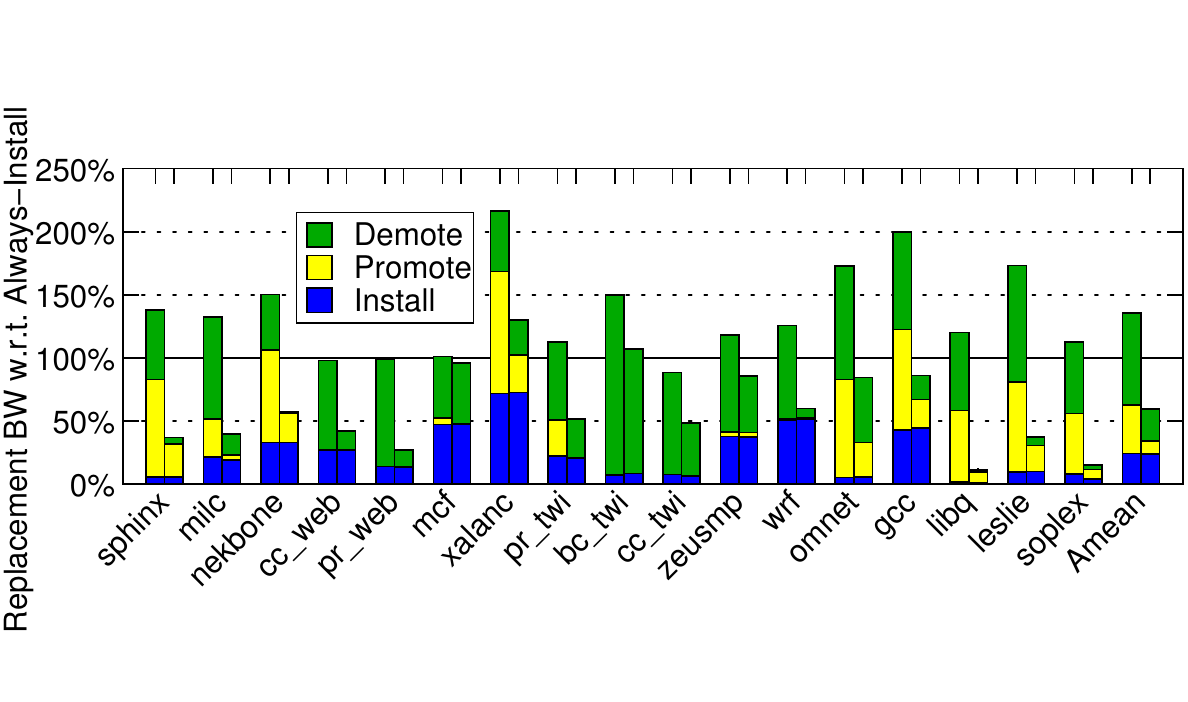}
		\vspace{-0.35in}
	\caption{Replacement and Install bandwidth consumption of base RRIP-AOB [left] and ETR on RRIP-AOB [right], normalized to Always-Install. ETR reduces 70\% of the bandwidth consumed in state update.}
	\label{fig:rrip_bw2} 
	\vspace{-0.05in}
\end{figure}

\subsection{Impact on Bandwidth}

ETR tries to reduce the bandwidth used for replacement state updates (RRPV promotion and demotion) to improve performance. To understand effectiveness of ETR, we divide the bandwidth used for cache replacement into three parts: installs, promotions, and demotions, and we normalize this consumption to the baseline design that uses bandwidth only for installs.  Figure~\ref{fig:rrip_bw2} shows the replacement bandwidth usage of base RRIP-AOB and ETR on RRIP-AOB, normalized to Always-Install. ETR saves 70\% of replacement state update bandwidth.  These bandwidth savings result in speedup.


\subsection{Impact on Performance}

Figure~\ref{fig:rrip_perf} shows the performance of RRIP-AOB, ETR on RRIP-AOB, and Ideal RRIP-AOB with no state update costs. ETR on RRIP-AOB bridges 70\% of the performance gap between RRIP-AOB and Ideal to achieve 18\% speedup, while incurring negligible SRAM storage costs.  Thus, our solutions of AOB and ETR make it practical to apply reuse-based policies to DRAM caches and get significant benefits while incurring negligible storage overheads.





\begin{figure*}[htb] 
	\vspace{-0.15in}
	\centering
	\includegraphics[height=1.5in]{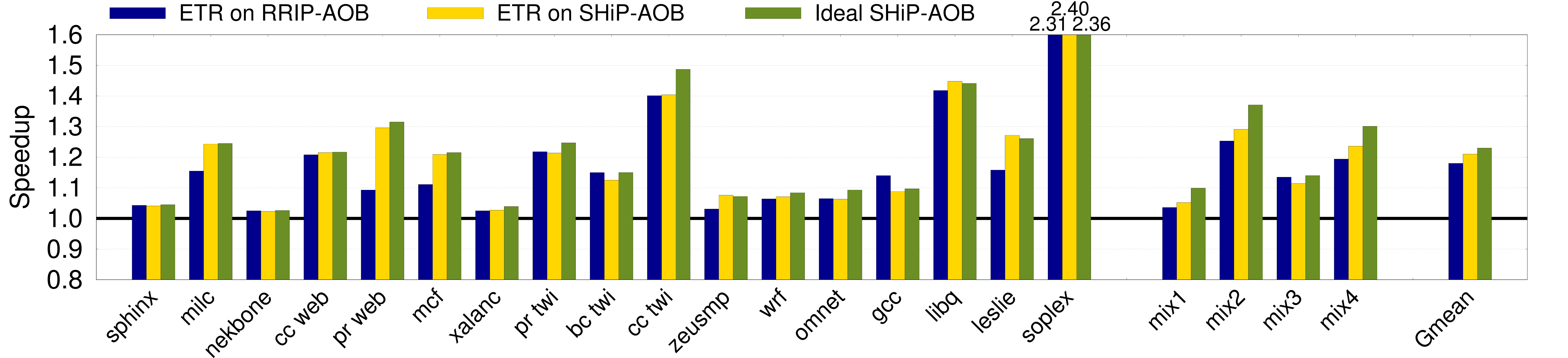}
	\vspace{-0.05in}
	\caption{Performance of ETR on RRIP-AOB, ETR on SHiP-AOB, and Ideal SHiP-AOB with no state update costs. }
	\label{fig:ship_perf} 
\end{figure*}

\section{Signature-Based Policies}

Thus far, we have discussed AOB and ETR only in the context of RRIP.  However, AOB and ETR are actually general techniques that enable formulating direct-mapped versions of replacement policies, as well as reducing the bandwidth needed to maintain replacement policy state. AOB and ETR can make even state-of-the-art signature-based policies~\cite{SHiP,SHiP++,Dead-block,Hawk-eye,Hawk-eye2} suitable for DRAM caches. We show how using {\em  Signature-based Hit Predictor (SHiP)}\cite{SHiP} as an example.


\begin{figure}[htb] 
	\centering
	\includegraphics[width=3 in]{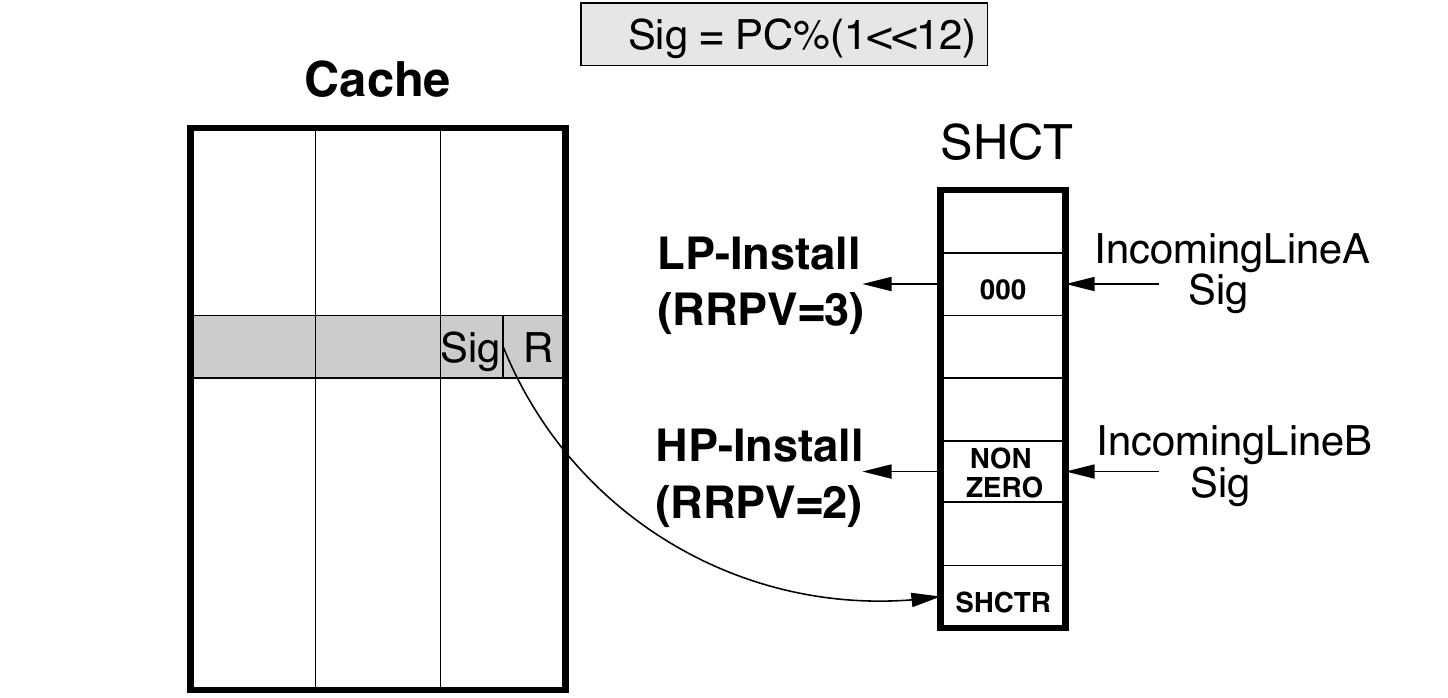}
	\vspace{-0.1in}
	\caption{Operation and Organization of SHiP.
		\normalsize}
	\label{fig:ship} 
	\vspace{-0.15in}
\end{figure}

\subsection{Operation of Conventional SHiP}

SHiP works by observing and learning which signatures correspond to low-reuse lines, and installing lines accessed by those signatures at low priority, as shown in Figure~\ref{fig:ship}. SHiP maintains signatures (PC) and reuse-bit (R-Bit) for tracking reuse of the signature. On eviction, the line increments or decrements a counter in the {\em Signature History Counter Table (SHCT)} based on the R-bit. On install, the SHCT decides if the incoming line is installed  with High-Priority (RRPV=2) or Low-Priority (RRPV=3), based on signature.

\subsection{Adapting SHiP to Direct-Mapped Cache}

Conventional SHiP design always installs the incoming line, either with High-Priority or Low-Priority. Unfortunately, with a direct-mapped cache, doing so will degenerate into the Always-Install policy (baseline). We extend SHiP in the context of direct-mapped caches using the option of bypassing with \textit{SHiP-AOB}. If the resident line has an RRPV=3, then the incoming line is always installed. If the resident line has an RRPV of less than 3, then we bypass the incoming line. However, we then demote the RRPV of the resident line {\em only} if the incoming line had a High-Priority install, and we skip the RRPV update for incoming lines with Low-Priority install. Thus, the advantage of SHiP-AOB is that it can reduce the bandwidth required to perform \textit{demotion} when the incoming line is predicted to have low reuse.

\subsection{Implementing SHiP for DRAM Cache}

To implement this bypassing version of SHiP for DRAM cache, we would need additional replacement metadata with each line: 12-bit signature + 1 R-Bit, in addition to the 2 bits for RRPV. Fortunately, these 15 bits of replacement metadata can still fit in the 18-20 unused bits available in ECC space of the KNL-Cache, so storage for the additional metadata for SHiP is not a concern.  The SHCT table still needs to be implemented in SRAM (similar to conventional SHiP); however, the SRAM overhead of the SHCT is only 1.5 kilobytes. 

Note that \ignore{the access for }updating the R-Bit occurs concurrently with the Promote operation (setting RRPV to 0), so no additional bandwidth is required for tracking the R-Bit. And, we force install 2\% of the time to get information on bypassed pages. For lines without a valid PC (e.g., writebacks), we train using a single PC-less SHCT entry.

\subsection{Impact on Bandwidth}

\begin{figure}[htb] 
	\vspace{-0.4in}
	\centering
	\includegraphics[width=1.0\columnwidth]{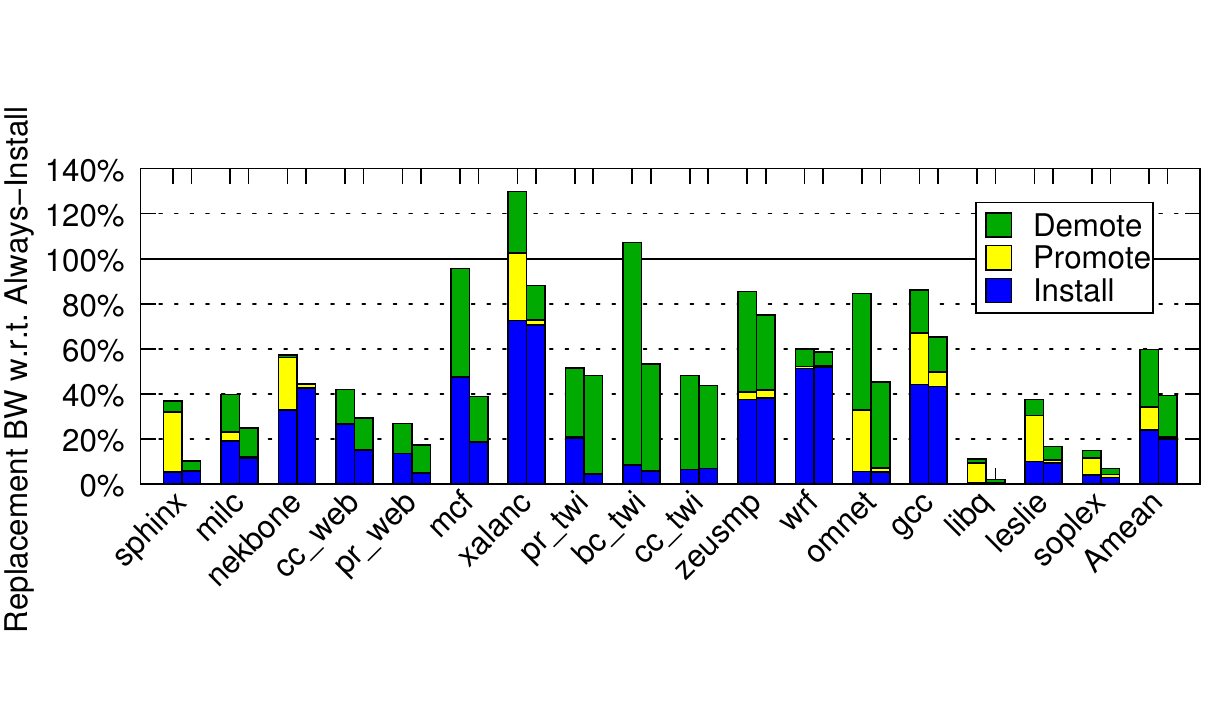}
	\vspace{-0.42in}
	\caption{Bandwidth usage of ETR on RRIP-AOB [left] and ETR on SHiP-AOB [right], normalized to Always-Install. SHiP-AOB further reduces BW for state update.}
	\label{fig:rrip_bw3} 
	\vspace{-0.12in}
\end{figure}

Figure~\ref{fig:rrip_bw3} shows the bandwidth breakdown (in terms of install, promotion, and demotion operations) for ETR on RRIP-AOB, and ETR on SHiP-AOB, normalized to the bandwidth consumed in doing installs for the always-install design. ETR on SHiP-AOB achieves further reduction in state update bandwidth compared to ETR on RRIP-AOB. ETR on SHiP-AOB is able to prevent the bandwidth for demotion operations by reducing update-cost when lines have predicted no-reuse, and this reduces the number of lines that need to be {\em promoted} back on hit. In particular,  ETR on SHiP-AOB benefits workloads that had poor spatial locality  (e.g., \textit{mcf}).

\subsection{Impact on Performance}

Figure~\ref{fig:ship_perf} shows the speedup of ETR on RRIP-AOB, ETR on SHiP-AOB, and the Idealized version of SHiP-AOB with no state update cost. Overall, ETR on SHiP-AOB achieves 21\% speedup, achieving most of the 23\% speedup with the Ideal design (which incurs impractical storage overheads).




\ignore{

\begin{figure}[htb] 
	\centering
\includegraphics[width=1.00\columnwidth]{GRAPHS/perf_ship_nowb.pdf}
	\vspace{-0.35in}
	\caption{Performance of SHiP. SHiP further reduces state update bandwidth to achieve 18\% speedup.}
	\label{fig:ship_perf} 
	\vspace{-0.05in}
\end{figure}
}

\ignore{

Scaling to 2-ways. (Not likely to go higher due to bandwidth costs).

{Parallel 2-way Cache}
Parallel 2-way. Read all ways on each access, determine hit/miss. Repl: Determine victim (tertiary: 2 ways or bypass). Stateless is fine. Stateful, 
QUESTION, where to store replacement state (as it has implications for bandwidth)? If you extend tag of each individually, updates to all counters (i.e. RRIP demotes) now need to write all ways. If you extend tag of only one way to store all counters, installing now needs to sometimes write to two locations. Let's look a state update bandwidth, with RRIP as an example

{RRIP-BYP on 2-way (needs trimming)}
RRIP-BYP, similar as before. If no victim, bypass and increment both counters. Works. But, state update costs changes depending on where state is stored. Should be optimized

    1-way state update: on Install (0), Promote (1), Demote-all (1)
    2-way together: on install (.5), promote (1), Demote-all (1)
    2-way individually: on install (0), promote (1), demote-all (2)
  ->2-way duplicated copies of together, with timestamp saying most recent together-state(can merge state update with install): Install (0), Promote (1), Demote-all (1)
    Explanations: 
        together: Only one way has replacement state. On install, might need to write to other location. 
        individually: RRIP updates multiple lines' state at a time (increment).
        Rather, duplicated copies: with timestamp saying whicch is momst recent
            // Both ways have state of both, with timestamp saying which one is the most recent. Merge state update with install.

    Bandwidth-efficient state management to maintain direct-mapped state update costs.

Effectiveness
    LRU,RRIP,RRIP-BYP+ETR. Better replacement, lower bandwidth.

    //Works, but has bad hit bandwidth that can hurt workloads that fit (sphinx). 
}

\section{Towards Set-Associative Designs}

We evaluate our solutions in the context of a direct-mapped cache, but our designs and insights can be made applicable to set-associative caches. 
A recent proposal ACCORD~\cite{ACCORD} tries to make DRAM caches set-associative, to improve hit rate albeit at an expense of bandwidth and latency~\cite{agarwal_ca,way_mru1,way_mru2}.  We compare with the recently proposed associative cache design, and show that ETR has higher potential as it improves both hit-rate \textit{and} bandwidth. Nonetheless, ETR on RRIP-AOB can be used in conjunction with set-associative designs for DRAM caches for even  better performance.

\begin{figure*}[htb] 
	\vspace{-0.15in}
	\centering
	\includegraphics[height=1.5 in]{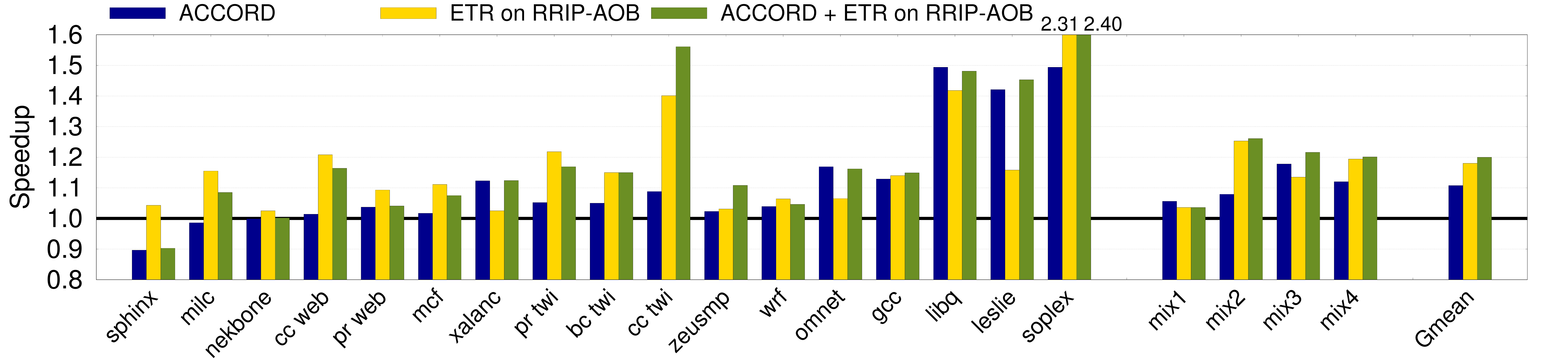}
	\vspace{-0.10in}
	\caption{Performance of set-associative ACCORD, ETR on RRIP-AOB, and ACCORD with ETR on RRIP-AOB. }
	
	\label{fig:accord_speedup} 
	\vspace{-0.13in}
\end{figure*}

\subsection{ACCORD: Predictable Associativity}

\begin{figure}[htb] 
	\vspace{-0.1in}
	\centering
	\includegraphics[width=1.0\columnwidth]{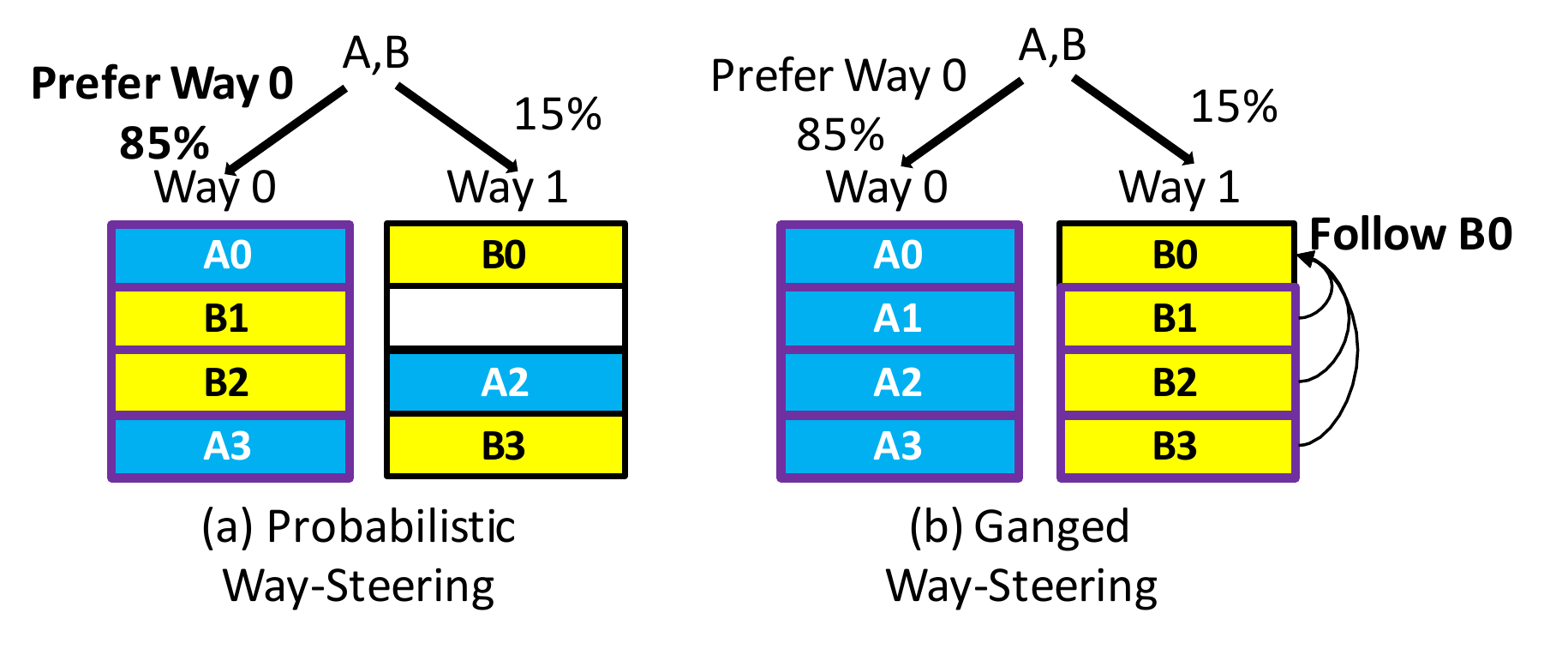}
	\vspace{-0.2in}
	\caption{ACCORD enables low-latency associativity for DRAM cache, by modifying install policy to make it easier to predict which way a line is in.}
	\label{fig:accord1} 
	\vspace{-0.1in}
\end{figure}

DRAM caches that store tag with data~\cite{Alloy,KNL} are known to be difficult to design for associativity--if there are multiple possible locations for a line, you may need several accesses to look up the possible locations for the line. A recent work, ACCORD~\cite{ACCORD}, proposes to modify replacement policy to make it easier to locate lines with simple way prediction methods. For example, it proposes Probabilistic Way-Steering (PWS) that biases install to a preferred way 85\% of the time based on address, as shown in Figure~\ref{fig:accord1}(a). It also proposes Ganged Way-Steering (GWS) that steers subsequent installs to a region into the same way, to enable per-region last-way-seen way-prediction to be accurate, shown in Figure~\ref{fig:accord1}(b).


Figure~\ref{fig:accord_speedup} and Figure~\ref{fig:accord_missrate} shows ACCORD provides 10\% speedup, as it reduces misses by 15\% but can cost extra bandwidth consumption (due to way-mispredictions and looking up multiple ways on miss). Meanwhile, ETR on RRIP-AOB provides a higher 18\% speedup, as it reduces misses by 10\% while simultaneously reducing DRAM cache bandwidth consumption. However, these ideas are not direct competitors. Associativity enables storing of multiple conflicting lines, and bypassing enables reducing install bandwidth while maintaining hit-rate. We design a solution that gets benefits of both.


\subsection{Combining ACCORD and Bypassing}

We want to obtain the hit-rate benefits associativity offers; however, we must do so while maintaining ACCORD's biased-install or we will face increased bandwidth cost to locate lines (due to frequent way-mispredictions).

\begin{figure}[htb] 
	\vspace{-0.13in}
	\centering
	\includegraphics[width=1.0\columnwidth]{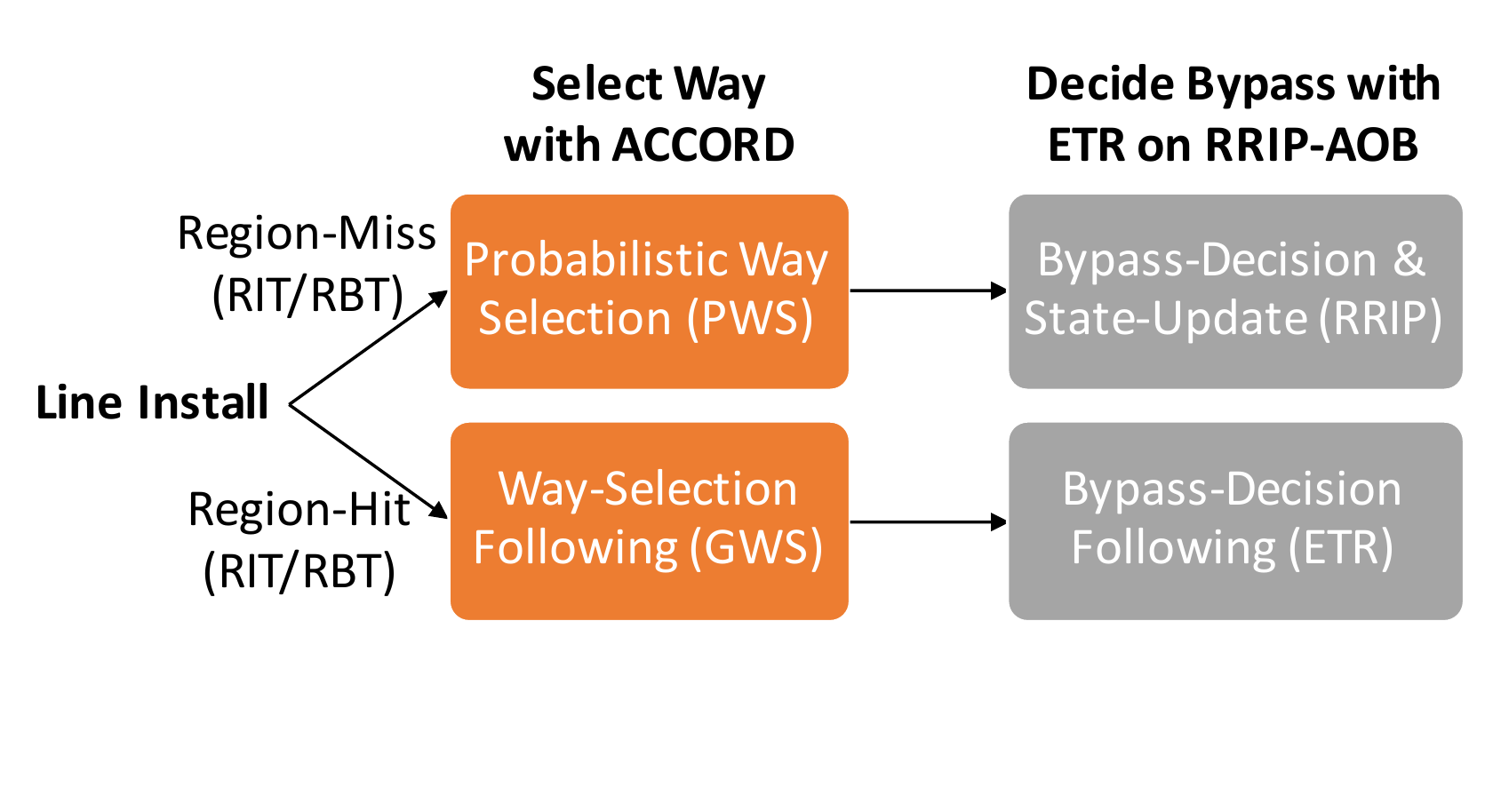}
	\vspace{-0.57in}
	\caption{We first use ACCORD to select which way to attempt install, then use RRIP-AOB to decide bypass.}
	\label{fig:accord2} 
	\vspace{-0.07in}
\end{figure}

To obtain benefits of both associativity and bypassing, we combine the two by viewing associativity as a {\em way-selection policy} and using a tiered decision. Figure~\ref{fig:accord2} shows our tiered decision tree: ACCORD is first used to select the way, and then ETR is used to determine bypass policy.

For the first install to a region (Region Miss in ACCORD-RIT and ETR-RBT), we use ACCORD PWS to select which way to attempt install. This will have a biased probability to install into a preferred-way, so we can maintain similar way prediction accuracy. We subsequently use RRIP-AOB to decide install or bypass into this particular way, and correspondingly update the state. This enables us to keep ACCORD's flexibility to use both ways and maintain high way prediction accuracy, as well as add RRIP's thrash-resistant replacement. This combination of the two enables high way prediction accuracy (ACCORD) and bandwidth-efficient state update and bypass (ETR on RRIP-AOB). 

\subsection{Effectiveness: Miss-rate and Speedup}

\begin{figure}[htb] 
	\vspace{-0.01in}
	\centering
	\includegraphics[width=1.0\columnwidth]{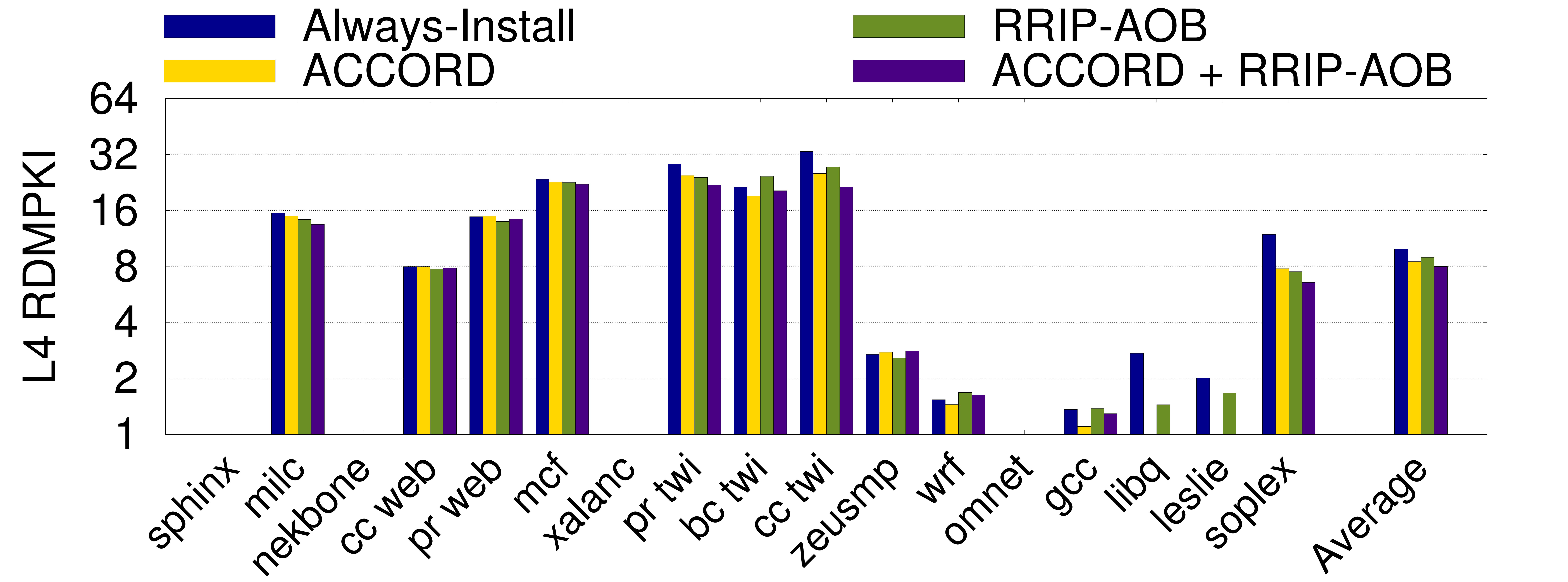}
	\vspace{-0.24in}
	\caption{L4 Read-Miss-Per-Kilo-Instruction of Always-Install, ACCORD, RRIP-AOB, and ACCORD + RRIP-AOB. Combination enables 20\% miss reduction.}
	\label{fig:accord_missrate} 
\end{figure}

Such a tiered decision process allows the bypass-policy ETR on RRIP-AOB to continue to obtain its thrash and scan-resistance, and bandwidth benefits. And, it enables ACCORD's associativity to utilize both ways to improve hit-rate, at minor cost to DRAM cache bandwidth. Figure~\ref{fig:accord_missrate} shows the read MPKI of ACCORD, \ignore{ETR on }RRIP-AOB, and the combination of the two. ACCORD reduces miss-rate by 15\%, \ignore{ETR on }RRIP-AOB reduces miss-rate by 10\%, and the combination of the two reduces miss-rate by 20\%. 

Figure~\ref{fig:accord_speedup} shows the performance of ACCORD, ETR on RRIP-AOB, and the combination of the two. ETR on RRIP-AOB improves both hit-rate and bandwidth, whereas ACCORD improves hit-rate at the cost of bandwidth. The combination of ACCORD and ETR on RRIP-AOB enables 20\% speedup, and show the concepts developed in ETR and RRIP-AOB are also applicable to set-associative DRAM cache designs. We discuss potential for improving other set-associative designs in Section~\ref{ssec:other_2way}.


\ignore{
\subsection{Comparison to Associativity}

We design ETR for a direct-mapped cache, but there can be a case for minimally-associative DRAM caches. KNL can implement 2-way cache by streaming all ways on each access, but, this design would be subject to bandwidth and latency overheads.
If such overheads can be avoided, Table~\ref{tab:associativity} shows an Ideal-2-way design can achieve 11.7\% speedup. Associative designs optimize for hit-rate, but come at a cost to bandwidth. In comparison, ETR obtains both hit-rate \textit{and} bandwidth benefits to achieve speedup of 19.4\%. Nonetheless, we can combine associativity and bypassing policies for further speedup.

\subsection{Design of ETR + Associativity}

To obtain benefits of both associativity and bypassing, we combine the two by viewing associativity as a {\em placement policy}. We combine them in two steps: first,  we use 2-way random-placement to decide way to attempt install, and second, we use ETR to decide bypass or install. 
Such a tiered decision process allows the bypass-policy (ETR) to continue to obtain its thrash and scan-resistance, and bandwidth benefits. And, it additionally enables associativity to eventually utilize both ways to improve hit-rate.
Table~\ref{tab:associativity} shows that combining bypassing policy and associativity enables 29.7\% speedup.

\begin{table}[hbt]
	\vspace{-.15 in}
	\centering
	\caption{Comparison to Associativity}
	\vspace{.05 in}
	\setlength{\tabcolsep}{4.3pt}{
	    \begin{tabular}{|c||c|c|c|}\hline
		 & Ideal-2-way & ETR 1-way & ETR + Ideal-2-way  \\ \hline \hline
		SPEC RATE  & +14.7\% & +18.9\% & +31.2\% \\ \hline
		SPEC MIX  & +12.6\% & +17.0\% & +27.0\% \\ \hline
		GAP  & +4.1\% & +22.8\% & +28.1\% \\ \hline \hline
		GMEAN26  & +11.7\% & +19.4\% & +29.7\% \\ \hline
	    \end{tabular}
    }
	\label{tab:associativity}
	\vspace{-.15 in}
\end{table}

}




\ignore{

Storage overhead. Tag, valid, dirty, tid, 10bits in l4. 12bit pc 3bit shctr needs 1.5KB sram. 1\% sets need 12bit pc, 1\% sets need 6bit virttag+valid.
256 threads needs 8bit tid, tag could be 5+2, dirty, valid, is 17bit. Pc could be truncated to 11. 8bit virttag+valid.

Stability of first-access to page. Histogram first-miss offset, per-workload

}

\newpage

\section{Results and Discussion}

In this section we present sensitivity studies and storage analysis. Due to space constraints, we limit these results to ETR implemented on RRIP-AOB.



\subsection{Multi-programmed Workloads}
\label{ssec:mixed}

\begin{figure}[htb] 
	\vspace{-0.11in}
	\centering
	\includegraphics[width=1.0\columnwidth]{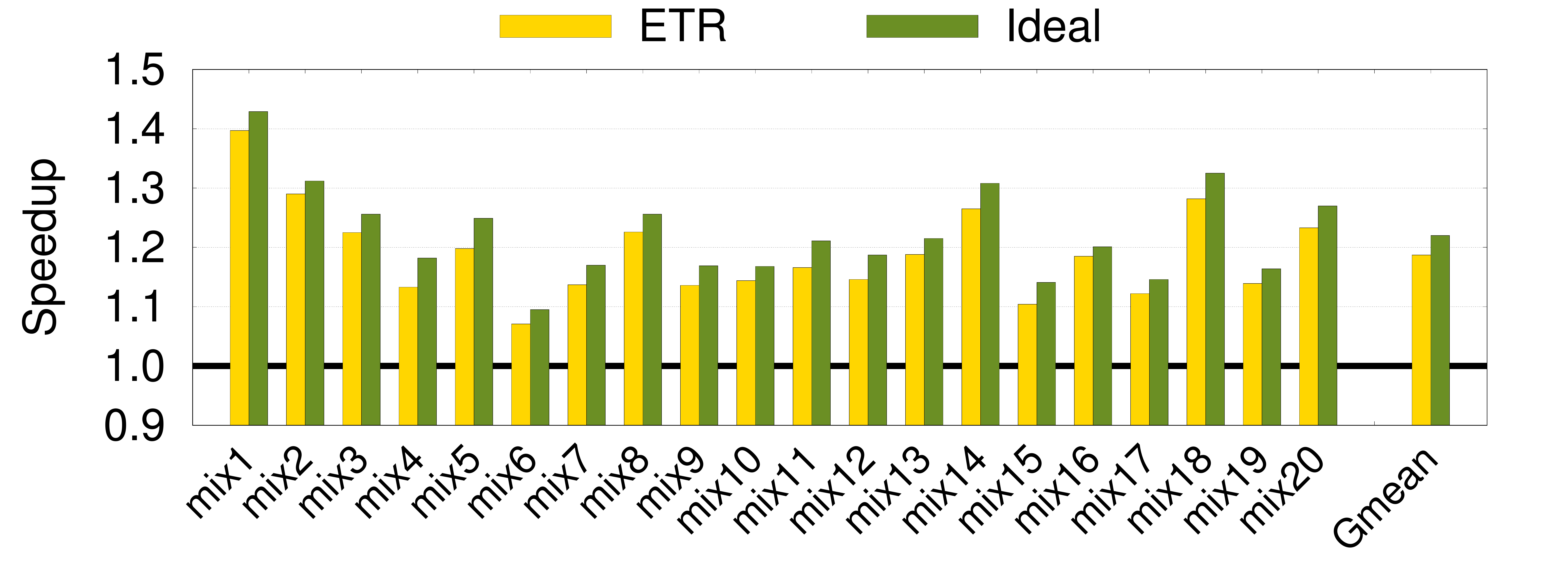}
	\vspace{-0.37in}
	\caption{Speedup of ETR on RRIP-AOB and Ideal RRIP-AOB on multi-programmed workloads.}
	\label{fig:mix_perf} 
	\vspace{-0.17in}
\end{figure}

To show robustness of our proposal to multi-programmed workloads, we evaluate over a larger set of 20 mix-application workloads. 
Figure~\ref{fig:mix_perf} shows that ETR provides 19\% speedup across 20 mixes, with no workloads experiencing slowdown.

\subsection{Impact on Energy and Power}

\begin{figure}[htb] 
	\vspace{-0.18in}
	\centering
	\includegraphics[width=1.00\columnwidth]{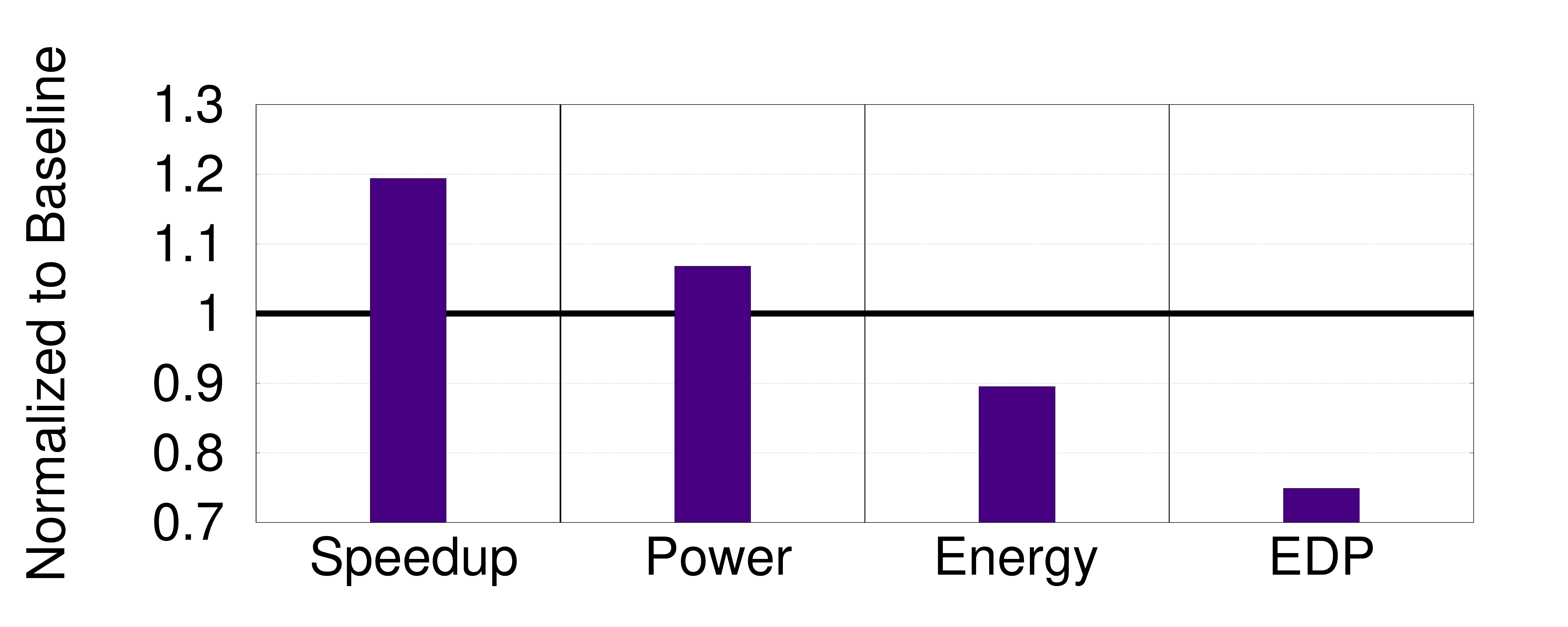}
	\vspace{-0.33in}
	\caption{Memory system energy of ETR on RRIP-AOB. Intelligent replacement reduces energy usage.}
	\label{fig:edp} 
	\vspace{-0.17in}
\end{figure}

Figure~\ref{fig:edp} shows DRAM cache + memory power, energy consumption, and
energy-delay-product (EDP) of a system using ETR, normalized to baseline DRAM cache. 
We model power and energy for stacked DRAM with\cite{Chandrasekar_date2013,Malladi_micro2012}, and model power and energy for non-volatile memory with\cite{lee:isca09}.
ETR reduces DRAM cache energy by reducing install and state update bandwidth, and provides lower main memory energy by improving DRAM cache hit-rate. Overall, ETR reduces energy consumption by 11\% and EDP by 24\%.

\subsection{Storage Requirements}

We analyze the SRAM storage overheads of ETR. ETR requires only a 128-entry 4B-per-entry Recent-Bypass Table, which needs 512B. Thus, our proposal can be easily built with negligible overheads within the memory controller. 

For DRAM storage overheads required for RRIP-AOB, we use the fact that the DRAM cache has 28 unused bits in the ECC, which can be used for tag and metadata (see Figure~\ref{fig:alloy}). Baseline uses 8-10 bits for tag, valid, and dirty bit. RRIP requires just 2 bits for RRPV. Tag-entry becomes 12 bits, which fit in 28 available bits.



\ignore{
\begin{table}[hbt]
	\centering
	\caption{SRAM Storage Requirements of ETR}
	\begin{tabular}{|c||c|}\hline
		ETR Component & Storage  \\ \hline \hline
		Recent-Bypass Table  & 512 Bytes \\ \hline
		ETR total & 512 Bytes \\ \hline
	\end{tabular}
	\label{fig:storage}
\end{table}
}

\subsection{Impact of Cache Size}

Table~\ref{tab:size} shows the speedup of ETR as the size of the
DRAM cache is varied from 1GB to 8GB. ETR on RRIP-AOB continues to provide
significant speedup across different cache sizes, ranging from 16.4\% at
1GB to 13.5\% at 8GB.  
As expected, when the cache size is increased, larger portions of the workload fit in, and there is reduced scope for improvement.

\begin{table}[hbt]
	\vspace{-.21 in}
	\centering
	\caption{ETR Sensitivity to Cache Size}
	\vspace{.03 in}

	\begin{tabular}{|c||c|}\hline
		Cache Size  & Avg. Speedup from ETR  \\ \hline \hline
		1.0GB   & 16.4\%  \\ \hline
		{\bf 2.0GB}   & {\bf 18.0\%} \\ \hline
		4.0GB   & 17.4\% \\ \hline
		8.0GB   & 13.5\% \\ \hline
	\end{tabular}
	\label{tab:size}
	\vspace{-.15 in}
\end{table}

\subsection{Impact of Memory Type}
\label{ssec:dram-memory}

We use a non-volatile main memory for our studies, but our benefits are not limited to NVM-backed systems only. We compare BEAR's Bandwidth Aware Bypass\cite{BEAR} with proposed ETR on DRAM-backed main memory in Table~\ref{tab:dram_memory}. ETR outperforms BEAR by intelligently bypassing lines and achieving better hit-rate and substantial bandwidth benefits.

\begin{table}[hbt]
	\vspace{-.14 in}
	\centering
	\vspace{-.05 in}
	\caption{ETR on DRAM-backed Memory}
	\vspace{.03 in}
	\setlength{\tabcolsep}{5.3pt}{
		\begin{tabular}{|c||c|c|}\hline
			& Bandwidth Aware Bypass & ETR  \\ \hline \hline
			SPEC RATE  & +7.4\% & +17.0\%  \\ \hline
			SPEC MIX  & +1.7\% & +16.0\%  \\ \hline
			GAP  & +4.8\% & +26.6\% \\ \hline \hline
			GMEAN26  & +5.7\% & +19.0\% \\ \hline
		\end{tabular}
	}
	\label{tab:dram_memory}
	\vspace{-.15 in}
\end{table}



\subsection{Impact of Region Size}
Table~\ref{tab:regionsize} shows the speedup of ETR as the region size is varied from 1KB to 4KB. Region size of 4KB (matching smallest OS page) provides best speedup of 18.0\% as it amortizes the most replacement-update costs.

\begin{table}[hbt]
	\vspace{-.15 in}
	\centering
	\caption{ETR Sensitivity to Region Size}
	\vspace{.03 in}

	\begin{tabular}{|c||c|}\hline
		Region Size  & Avg. Speedup from ETR  \\ \hline \hline
		1KB   & 17.8\% \\ \hline
		2KB   & 18.0\% \\ \hline
		{\bf 4KB}   & {\bf 18.0\%} \\ \hline

	\end{tabular}
	
	\label{tab:regionsize}
	\vspace{-.15 in}
\end{table}

\subsection{Impact on Other 2-Way Designs}
\label{ssec:other_2way}

We evaluate our design on the state-of-the-art set-associative DRAM cache design ACCORD~\cite{ACCORD}, but the intelligent replacement offered by RRIP-AOB can be useful for other set-associative designs as well. 
To isolate the impact of intelligent replacement (without considering bandwidth costs specific to each DRAM cache organization), Table~\ref{tab:other_repl} shows the impact on average L4 misses when using different replacement policies for 1-way and 2-way caches, normalized to the baseline 1-way always-install policy. As expected, our RRIP-AOB achieves the highest miss reduction for 1-way caches as it enables intelligent reuse-based replacement policy for 1-way caches. Additionally, RRIP-AOB also achieves the highest miss reduction for 2-way caches. This is because RRIP-AOB can intelligently decide to bypass in the case the cache set is storing multiple useful lines.

\begin{table}[!h]
	\vspace{-.15 in}
	\centering
	\caption{RRIP-AOB Impact on Misses for 1-2 Way L4}
	\vspace{.03 in}
	\resizebox{3.275in}{!}{
    \setlength{\tabcolsep}{2.3pt}{
	\begin{tabular}{|c||c|}\hline
		Replacement Policy  & Impact on Avg. L4 Misses \\ \hline \hline
		1-way Always-Install   & -0.0\%  \\ \hline
		1-way Probabilistic Bypass~\cite{BEAR}   & -1.6\%  \\ \hline
		{\bf 1-way RRIP-AOB}   & {\bf -10.4\%}  \\ \hline
		2-way Random   & -14.6\%  \\ \hline
		2-way LRU   &  -15.5\% \\ \hline
		2-way RRIP   & -19.7\% \\ \hline
		{\bf 2-way RRIP-AOB}   & {\bf -26.6\%} \\ \hline
	\end{tabular}
	}
	}
	\label{tab:other_repl}
	\vspace{-.2 in}
\end{table}

\begin{figure*}[htb] 
	\vspace{-0.31in}
	\centering
	\includegraphics[height=1.5in]{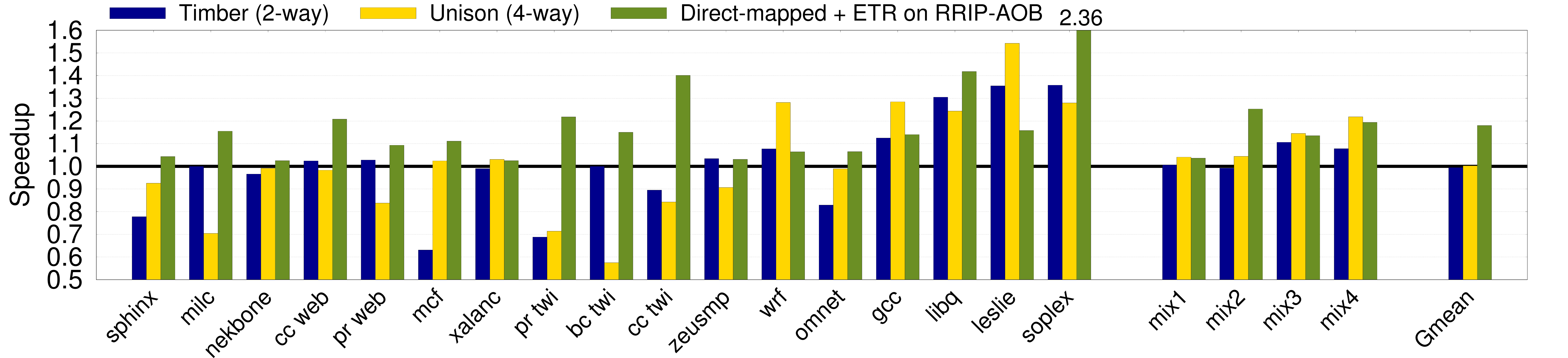}
	\vspace{-0.2in}
	\caption{Speedup of line-based~\cite{timber} and page-based~\cite{unison} set-associative DRAM caches, rel. to baseline direct-mapped DRAM cache~\cite{KNL,Alloy}. Proposed ETR on RRIP-AOB enables direct-mapped DRAM caches to obtain the hit-rate benefits of intelligent cache replacement, without needing to pay additional bandwidth to maintain set-associative tags.}

	\label{fig:timber} 
	\vspace{-0.15in}
\end{figure*}

\newpage



\ignore{
\textit{Quality-of-Service:} However, purely optimizing for hit-rate and bandwidth under a shared cache can cause less-memory-intensive workloads to have difficulty obtaining space in the cache. In such mixed-application scenarios, we would like to obtain some level of Quality-of-Service(QoS) guarantees, where each core can achieve similar level of performance as if it had a \textit{partitioned} slice of the DRAM cache. 

To accomplish QoS, we estimate hit-rate under partitioned cache by sampling 1\% sets and installing virtual tags into DRAM-ECC space under the partitioned indexing. We then compare estimated individual partitioned-cache hit-rate with current individual shared-cache hit-rate. If a core sees significantly degraded hit-rate (\textgreater5\%) under shared, we protect cache space for that core. By protecting degraded cores, we are able to ensure individual core performance to be at-worst similar to performance under a partitioned cache.

\begin{figure}[htb] 
	\vspace{-0.15in}
	\centering
\includegraphics[width=1.00\columnwidth]{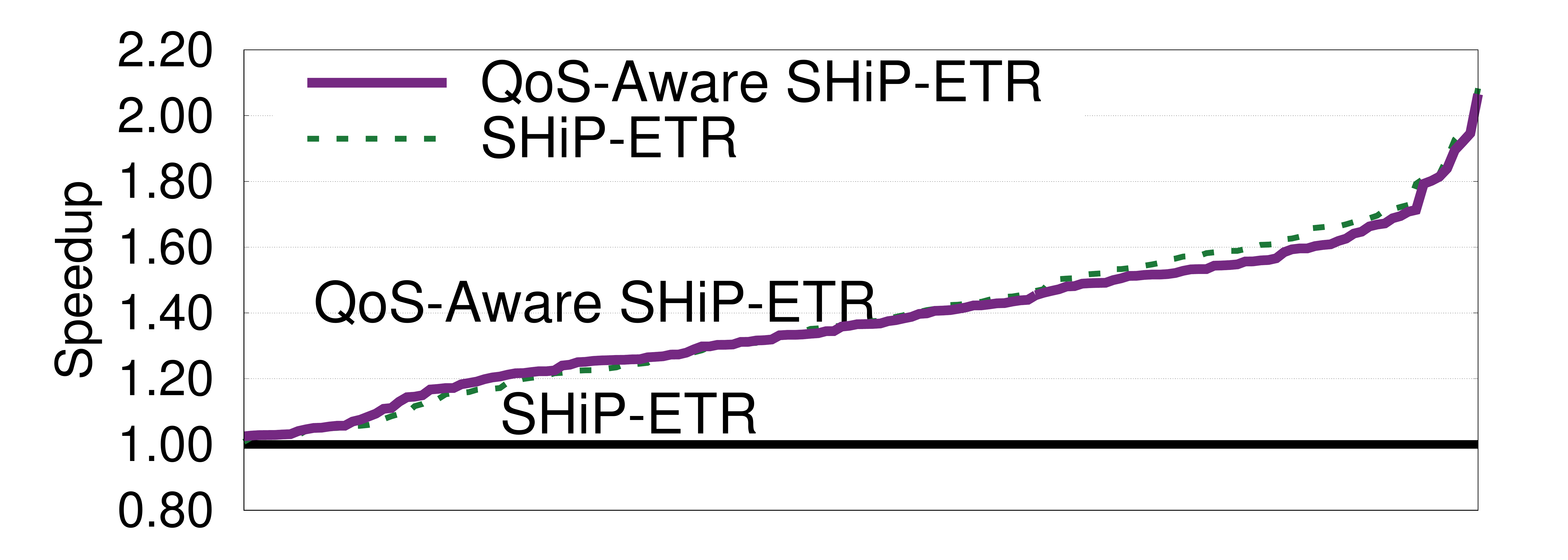}
	\vspace{-0.25in}
	\caption{Individual workload performance of SHiP-ETR and QoS-Aware SHiP-ETR on 20 8-thread Mixed-Application workloads, sorted by speedup. QoS-Aware SHiP-ETR ensures better worst-case performance.}
	\label{fig:qos_scurve} 
	\vspace{-0.15in}
\end{figure}

Figure~\ref{fig:qos_scurve} shows the individual speedup of each application (8*20=160 total) in the mixes under QoS-unaware and QoS-aware SHiP-ETR, relative to individual performance under partitioned DRAM cache. Each individual application sees speedup relative to the partitioned case due to effective shared cache management under our SHiP-ETR. With our QoS-aware policy, we are able to additionally protect lower-MPKI workloads against high-MPKI, large-footprint workloads to ensure better worst-case performance.
}
\section{Related Work}
\subsection{Replacement / Bypassing policies}


\textit{Recency}-based replacement policies\cite{lru2,lru1,lru3} install incoming lines at highest priority, which degenerate into always-install baseline. \textit{Probabilistic} replacement policies\cite{DIP,TADIP}, become probabilistic bypass\cite{crc1} in Figure~\ref{fig:bab}.
\textit{Frequency}-based replacement \cite{freq1,VWAY,freq2,freq4,freq5} or \textit{Reuse}-based replacement\cite{RRIP,rrip2,rrip3,crc1,ctrbypass} try to predict and keep most-frequently used lines in the cache. We design a bypassing version of RRIP, RRIP-AOB, and implement ETR on our RRIP-AOB as an example of this class of policies, but our ETR scheme can be easily used to reduce update-cost of other frequency and reuse-based replacement algorithms. \textit{Signature}-based replacement\cite{SHiP,SHiP++,Dead-block,Hawk-eye,Hawk-eye2} attempt to predict line reuse based on signatures (e.g., PC). We develop a bypassing version of SHiP, called SHiP-AOB, to show how to implement signature-based replacement on caches with low associativity.



\subsection{Line-based DRAM Caches}

\label{ssec:line}

In our study, we use the DRAM cache organization used in Intel's Knights-Landing~\cite{KNL} that is direct-mapped and stores each tag next to its data as our baseline. This organization is the commercial implementation of many research efforts that store Tag-With-Data\cite{BEAR,Alloy,candy} to improve latency and reduce bandwidth consumption. We compare with recent enhancements in Figure~\ref{fig:bab} (90\%-Bypass and BAB\cite{BEAR}).

Alternative designs such as Sim et al.\cite{simcache} take a different approach to storing tags via {\em tag grouping}. For such caches, a tag-only line is placed along with data in the same row buffer\cite{LHCache,simcache,timber,atcache,bwp}. Such caches require separate accesses for tag and data, which can cost significant bandwidth and latency. Timber is an enhancement that proposes to mitigate tag lookup by using a tag-cache and exploiting spatial locality (by co-locating tags and metadata from multiple sets)\cite{timber}. We compare with Timber as a representative of the grouped tag / metadata approach in Figure~\ref{fig:timber}. Such approaches can enable associativity, but pay substantial bandwidth to access and update tags when the tag cache has poor hit-rate, due to large footprint and poor spatial locality (e.g., \textit{mcf} and \textit{pr twi}). Our ETR on RRIP-AOB, on the other hand, enables intelligent replacement without needing to access tags separately, and outperforms such tag-grouped approaches.


\subsection{Page-based DRAM Caches}

\label{ssec:page}

An alternate approach to designing DRAM caches is to use large granularity caches to reduce tag and metadata overhead, in hardware\cite{FootprintCache:ISCA2013:MICRO2014papers,unison} or software\cite{tagless,tagless2,loh_freq,banshee}. The reduction in tag requirements enable more space for associativity and replacement metadata. Such large-granularity caches often employ recency-based replacement\cite{unison,tagless,tagless2} or frequency-based replacement\cite{loh_freq,banshee} that would otherwise be too expensive in line-granularity caches. We compare with Unison cache\cite{unison} (hardware-managed, 4-way, page-based sectored cache with LRU replacement, separate tag lookup) as a representative of page-based designs, in Figure~\ref{fig:timber}. The associativity and replacement Unison offers enable it to frequently outperform the baseline DRAM cache. However, the large linesize of Unison often limits it from using a large portion of the cache (e.g., \textit{pr twi} and \textit{bc twi}), and, the separate tag and data lookup often wastes significant bandwidth. 
Our ETR on RRIP-AOB, on the other hand, enables direct-mapped DRAM caches to obtain \textit{intelligent replacement} without sacrificing \textit{cache-utilization} or \textit{bandwidth-efficiency}, to outperform such page-based approaches.


\ignore{
LRU, BIP/DIP/BEAR-BAB, SRRIP, SHiP(SHiP,SHiP++,Deadblock,Multiperspective,Bellady).
We take in insights of all prior policies and apply it to improve hit-rate of DRAM caches. However, we these policies require per-line state. Scaled to DRAM cache, State-in-SRAM is storage-expensive, state-in-DRAM is bw-expensive--we solve this problem. In addition, we need special consideration for writebacks.

Previous solutions commonly implement writeback (it is okay writeback when you have sufficient ways). However, implementing writeback for direct-mapped caches can waste significant cache space when lines are not reused. Write-around is not new, but dynamic write-back / write-around is important for maintaining performance in multi-core scenario.
}



\ignore{
Our work utilizes the insights developed from effective last-level cache management policies, to develop a cheap and effective DRAM cache bypassing policy. Frequency-friendly replacement policies\cite{freq1,freq2,VWAY,freq4,freq5} try to predict and keep most-frequently used lines in the cache. 
Recency-friendly replacement policies\cite{lru1,lru2,lru3} work by installing incoming lines with the highest priority and promoting them on reuse, with the insight that recently used lines are more likely to be re-used. However, when the working set is larger than the number of ways can support, thrashing (new lines continuously evict older useful lines) can often occur and give low hit rate. In such cases, Dynamic Insertion Policy\cite{DIP} or Bandwidth-Aware Bypass\cite{BEAR} can dynamically choose to protect working set by mostly installing at lowest-priority.
However, such global policies are ineffective for large caches (see Figure~\ref{fig:bab}) as they are too coarse-grain and are not able to protect against fine-grain thrashing. Re-Reference Interval Predictor\cite{RRIP} and other reuse-distance based works\cite{rrip2,rrip3} are thrash-and-scan-resistant per-set replacement policies often used in last-level caches. RRIP works on the basis that re-used lines are most likely to be re-used, and most other lines can be bypassed by default. We can further add PC-awareness with SHiP or Dead-block prediction to improve the install decisions\cite{SHIP,SHIP++,Dead-block,Hawk-eye}. Write-induced interference has also been observed in several works\cite{write1,qureshi:hpca10,virtual_write_queue,last-write_prediction}.
Our RRIP-ETR and SHiP-ETR incorporates these state-of-art replacement policies, and further tunes them for bandwidth-efficiency by COORDinating replacement policy across sets.
In addition, we show how effectively handle write-back or write-no-allocate (write-around) for direct-mapped caches with Dynamic Write-Back, to optimize both main-memory write-traffic and cache-utilization under a direct-mapped cache.
}



\ignore{
Prior work on replacement policies typically work by protecting a portion of the working set\cite{DIP,TADIP,RRIP}, or determining which lines are less likely to be used and installing those lines at lower priority~\cite{SHiP,SHiP++}. Recency-friendly replacement works by installing incoming lines with the highest priority and promoting them on reuse, with the insight that recently used lines are more likely to be re-used. However, when the working set is larger than the number of ways can support, thrashing (new lines continuously evict older useful lines) can often occur and give low hit rate. Bimodal Insertion Policy (BIP) and Dynamic Insertion Policy (DIP)\cite{DIP} are global replacement policies that target this weakness and propose to install most incoming lines at lowest priority, to enable the cache to protect much of its working set. For example, BEAR's Bandwidth-Aware-Bypass in Figure~\ref{fig:bab} is a bypassing formulation of DIP that enables working-set protection. However, global policies are often too coarse-grain and are not able to protect against fine-grain thrashing. 

Re-Reference Interval Predictor\cite{RRIP} is a thrash-and-scan-resistant per-set replacement policy often used in last-level caches. RRIP works on the basis that re-used lines are most likely to be re-used, and most other lines can be bypassed by default. RRIP works by installing lines at low priority, and saturating to highest priority upon first re-use. This policy has found to be effective for last-level caches and should work well for DRAM caches. However, when scaled to DRAM caches, RRIP needs significant per-line storage for state (3-bit state would need 12MB SRAM to store on-chip), and significant state update on hit and state update on victim selection (bandwidth-inefficient to store in DRAM).

A follow-up work Signature-based Hit Predictor (SHiP)~\cite{SHiP,SHiP++} suggests that we can provide fine-granularity working-set protection by installing lines that are unlikely-to-be-re-used at lowest priority. SHiP works by first observing install-signatures (the PC that caused the miss) and re-use information (if the line is re-used during its lifetime), and then learning which signatures correspond with re-use. SHiP learns which PC's are unlikely to re-use lines and installs the lines requested from those PC's at lowest priority. 
}

\ignore{
We use line-based KNL-cache. TAD similar to Alloy/BEAR/CANDY.  BEAR is most similar--Bypass-90\% to reduce install bandwidth. But too coarse, only 3\% speedup.

Grouped-tags. But cost significant bandwidth to read grouped tags.
}

\ignore{
In our study, we use the DRAM cache organization used in Intel's Knights-Landing (KNL-Cache) that associates tags with each data line. This Tags-with-Data style cache has been used in many previous studies\cite{Alloy,BEAR,candy}. Of notice, BEAR\cite{BEAR} attempts to reduce DRAM cache install-bandwidth by bypassing the cache 90\% probabilistically when doing so would not impact hit-rate. BEAR's Bandwidth-Aware-Bypass does provide minor speedup (3\% in Figure~\ref{fig:bab}), but it is too coarse-grain and can miss out on many opportunities to bypass without affecting hit-rate. Meanwhile, our SHiP-ETR avoids 80\% of the install bandwidth while maintaining higher hit-rate to achieve 21\% speedup.

Other studies take different approaches to storing tags via {\em tag grouping}.
For such caches, a tag-only line is placed along with data in the same row buffer\cite{LHCache,simcache,timber,atcache,bwp}.
For example, Aggressive Tag-Cache\cite{atcache} groups tags of a
15-way set into a single cache line and accesses it separately from
data, and, Timber\cite{timber} additionally uses a small structure that keeps 
recently accessed tag lines. Such caches depend on spatial locality to amortize tag lookup cost, and can perform poorly for workloads with poor spatial locality. Nonetheless, our bypassing policies can still be used to improve hit-rate for such caches.
}

\ignore{
recency
	-> always-install
frequency / reuse distance
	-> choose RRIP as representative to implement
signature based
    -> choose SHiP as representative to implement

write aware
    group writes for write locality
    	-> ETR grouped installs causes writebacks to have locality
    last-write prediction (signature-based)
        -> signatures for writes require substantial storage throughout cache hierarchy. simple scheme works. in addition, have to be more proactive (by write-no-allocate).
}


\ignore{
Unison cache. Sectored cache with 4-ways, with LRU replacement. Banshee/Tag tables(tagless) uses page-table/TLB with 4-ways, with frequency-based replacement. Reduce storage overhead by maintaining state per page. However, poor cache utilization for non-spatial workloads (mcf, graph) that preclude benefits, and coarse-grain replacement information (whereas SHiP-ETR still works per individual lines). Nonetheless, our SHiP can be applied on top of such caches to provide improved hit-rate.
}
\ignore{

An alternate approach to designing DRAM caches is use sectored caches\cite{FootprintCache:ISCA2013:MICRO2014papers,unison} and page-granularity caches\cite{tagless,tagless2,banshee,loh_freq} to reduce the storage required for the tag-store. Such caches use LRU or frequency-based replacement policies to keep the best pages in the cache. However, their large line size can prevent large amounts of the cache from being used when workloads exhibit poor spatial locality. Nonetheless, RRIP-ETR and SHiP-ETR can be applied on top of these caches to enable improved page replacement and bandwidth consumption.

}
	
\ignore{

intelligent replacement for practical DRAM caches, such as those in KNL. 
Such caches are direct-mapped for efficient tag lookup. 
However, low hit-rate.
Can use intelligent replacement policy to decide if incoming line should be bypassed.
However, intelligent policies typically require tracking reuse, which needs update on hits and bypass. 
These bandwidth costs limit the performance that can be gained. We

DRAM caches are direct-mapped, and co-locate tag with data within DRAM and stream out both in one access.
Need effective way to improve hit-rate.

}


\vspace{-.05in}
\section{Conclusion}
\ignore{
This paper investigates improving hit-rate for direct-mapped DRAM caches by utilizing reuse-based replacement polices. 
We would like to use the most effective replacement policies to improve DRAM cache hit-rate. 
Unfortunately, state-of-the-art policies based on reuse 
information are designed to compare multiple counter values within the set and then make the decision of a replacement victim. As such, these policies become ill-defined and inapplicable for direct mapped cache (they degenerate into either always-install or always-bypass policies). 
To make reuse-based policies, such as RRIP, applicable to direct-mapped DRAM caches, we propose a bypass formulation of RRIP called \textit{RRIP Age-On-Bypass (RRIP-AOB)}. RRIP-AOB leverages the insight that the event of bypassing in a direct-mapped cache, should also age the reuse state of the resident line.
}

This paper investigates improving hit-rate for direct-mapped DRAM caches by utilizing reuse-based replacement polices. 
We would like to use the most effective replacement policies to improve DRAM cache hit-rate. 
Unfortunately, state-of-the-art policies based on reuse are designed to compare multiple counter values within the set to decide a  replacement victim. As such, these policies become ill-defined and inapplicable for direct mapped caches (they degenerate into either always-install or always-bypass policies). 
To make reuse-based policies, such as RRIP, applicable to direct-mapped DRAM caches, we propose a bypass formulation of RRIP called \textit{RRIP Age-On-Bypass (RRIP-AOB)}. RRIP-AOB leverages the insight that the event of bypassing in a direct-mapped cache, should also age the reuse state of the resident line.


Similar to RRIP, RRIP-AOB needs per-line reuse counters. Maintaining such reuse state in DRAM costs significant bandwidth (\textit{promote} on hit, \textit{demote} on bypass). We investigate methods to reduce bandwidth overhead of keeping reuse state in DRAM. We propose \textit{Efficient Tracking of Reuse (ETR)} to reduce state update costs. ETR builds upon an observation that, at any given time, many lines from a 4KB region are \textit{coresident} and \textit{have similar reuse state}. If we select a representative (e.g., first-conflicting set in a region) and maintain accurate reuse state for just that line, we can use that line's reuse to infer rest of the lines' reuse without update-cost. ETR reduces the bandwidth for tracking replacement state by 70\% while maintaining similar hit-rate. 
Our evaluations with a 2GB DRAM cache, show that ETR on RRIP-AOB provides a speedup of 18.0\% while incurring an SRAM cost of \textless 1KB. ETR on RRIP-AOB performs within 2\% of an idealized design that does not incur bandwidth for state update.




\ignore{
Current DRAM caches are \textit{direct-mapped}, with \textit{tags stored in ECC bits}, to handle the issue of tag-storage and tag-access. Such a direct-mapped design is effective for reducing access-latency and bandwidth consumption. However, this direct-mapped organization comes at a cost to hit-rate. In this work, we target improving DRAM cache hit-rate.

A traditional method for improving cache hit-rate is using effective replacement policies. However, intelligent replacement policies (e.g., RRIP) typically require remembering reuse. We can store reuse and replacement state alongside tags in the DRAM-ECC space. But, maintaining this Replacement-State in DRAM (RSD) incurs significant bandwidth overhead. For example, Re-Reference Interval Predictor requires \textit{promote} bandwidth on hit, and \textit{demote} bandwidth when selecting victims. These bandwidth overheads limit performance potential of the RSD approach to 10\% speedup. We can store Replacement-State in SRAM (RSS) to avoid these bandwidth costs to achieve 21\% speedup. However, storing replacement state for gigascale caches on-chip requires impractically large multi-MegaByte SRAM storage (8MB).

To tackle SRAM storage and DRAM bandwidth costs, we implement a COORDinated-RSD-RRIP by storing replacement state in DRAM ECC and reducing state update bandwidth by coordinating replacement policy across sets. COORD-RSD-RRIP is based on the observation that replacement and eviction are often done at a page-granularity--if we maintain state for a representative (first conflicting set in a page), we can maintain good replacement state for the rest of the lines in the page. This enables reduction of state updates to 1/64 for workloads with good spatial locality. Coordinating replacement policy with RSD-COORD-RRIP reduces 70\% of the update costs as we now only need to maintain replacement information for just the first conflicting set of a page. This coordination improves speedup from 10\% to 16\%.

However, some write-intensive workloads can have poor performance (-15\%) compared to Always-Install, so, we find that we need make the replacement policy write-aware. Our RSD-COORD-RRIP treats demand requests and write requests identically. This results in write requests often bypassing the cache, which is a problem because we end up directly writing to write-sensitive non-volatile main memory. In the case writes have no reuse, such a default write-no-allocate policy optimizes for cache utilization. However, in the case there is reuse or rewriting to the line, a writeback cache (always-install writes) would be able to reduce write traffic. To dynamically choose between write-no-allocate and writeback, we propose a dynamic writeback/write-no-allocate policy based on reuse characteristics of written lines. 
Such a dynamic write-aware policy improves speedup from 16.0\% to 19.4\% speedup across our detailed workloads, with no workloads experiencing slowdown.

And, if we had PC-information, we can further reduce state update costs by making the policy PC-aware with SHiP. If SHiP predicts an incoming line has no reuse, we can often directly bypass and avoid state update for the resident line. To observe reuse, we store installing-PC bits and reuse-bit in spare bits of DRAM-ECC and update accordingly on hit and eviction. This can be done without significant SRAM-storage (stored in DRAM-ECC) and DRAM-bandwidth (RRIP already needs update-on-hit). In total, our bandwidth-efficient intelligent (Sig-ETR + DWP) replacement policy is able to achieve 21\% speedup and 26\% reduction in energy-delay-product in an NVM-backed system by improving hit-rate, reducing state update and install bandwidth, and writing-back when appropriate. 
}

\ignore{
As Practical DRAM caches (KNL-Cache) store tag-with-data, they are likely to be minimally-associative or direct-mapped to avoid bandwidth and latency costs. Our goal is to improve hit-rate for such minimally-associative caches. We can improve hit-rate for direct-mapped caches by formulating last-level cache replacement policies into DRAM cache bypassing policies. However, intelligent replacement policies typically require significant per-line storage and frequent state update. When applied to a gigascale DRAM cache, replacement status bits can take multi-MegaBytes of storage. Storing state-in-SRAM would be impractical as it the state required would be similar in size to last-level cache (8MB). And storing state-in-DRAM takes significant bandwidth to update (update-on-hit, update-on-victim-selection). An effective DRAM cache replacement policy should additionally minimize both SRAM-storage and DRAM-bandwidth costs.

To avoid significant SRAM storage cost, we propose implementing RRIP and storing state-in-DRAM (in spare bits of ECC). SRAM-RRIP with an impractical zero-latency state-in-SRAM can achieve up to 21\% speedup. Unfortunately, DRAM-RRIP that stores state-in-DRAM requires significant bandwidth to maintain state and can achieve only 10\% speedup. To mitigate state update costs, we propose coordinating bypassing policy across sets with TRIM-RRIP. TRIM-RRIP maintains state for the first-conflicting set in a page and forces subsequent accesses to the page to follow the same bypass decision. TRIM-RRIP maintains the same hit-rate as DRAM-RRIP while reducing the state update costs by 70\% (up to 98\% for streaming workloads). TRIM-RRIP achieves 16\% speedup while costing \textless512B of SRAM storage.

We can further reduce state update costs by making the policy PC-aware with SHiP. If SHiP predicts an incoming line has no reuse, we can often directly bypass and avoid state update for the resident line. To implement the observation phase, we store installing-PC-bits and reuse-bit in spare bits of DRAM-ECC and update accordingly on hit and eviction. This can be done without SRAM-storage (stored in DRAM-ECC) and DRAM-bandwidth (RRIP already needs update-on-hit). For the learning phase, we need to store signature reuse or no-reuse in a small on-chip Signature History Counter Table that requires 1.5KB. DRAM-SHiP enables further reduction in state update cost. 

However, some write-intensive workloads can degrade performance under mixed-application runs, so we must make the replacement policy write-aware. Our default DRAM-RRIP treats demand requests and write requests identically. This results in write requests often bypassing the cache, and directly writing to write-sensitive non-volatile main memory. In the case writes have no reuse, such a default write-no-allocate policy optimizes for cache utilization. However, in the case there is reuse or rewriting to the line, a writeback cache (always-install writes) would be able to reduce write traffic. To dynamically choose between write-no-allocate and writeback. we propose a write-aware enhancement to SHiP. 
Such a dynamic write-aware policy enables 21\% speedup across our detailed workloads and a robust 21\% speedup across mixed-application runs, with no workloads experiencing slowdown.

All in all, DRAM-SHiP enables significant hit-rate and bandwidth benefits for a Knights Landing DRAM cache at minimal SRAM storage cost(\textless 2KB) through state-in-DRAM and bandwidth-efficient design. We believe DRAM-SHiP's low storage overhead and minimal complexity over existing designs will make it appealing for industry adoption.
}

\bibliographystyle{ieeetr}

\bibliography{references}

\end{document}